\documentclass[twocolumn,numberedappendix,iop]{openjournal}
\usepackage{graphicx,amsmath,amssymb,amstext}
\usepackage{amsbsy,amsfonts,amsthm,color}
\usepackage[colorlinks,linkcolor=blue,citecolor=blue,urlcolor=blue ]{hyperref}
\usepackage[utf8]{inputenc}
\usepackage{float}
\usepackage{placeins} 
\usepackage{xcolor}
\usepackage{ulem}
\usepackage[T1]{fontenc}
\usepackage[title]{appendix}
\usepackage{array}
\usepackage{multirow}
\usepackage{comment}

\begin{document}

\title{Testing gravitational physics by combining DESI DR1 and weak lensing \\ \vspace{1mm} datasets using the $E_G$ estimator \vspace{-4em}}

\author{
S.~J.~Rauhut,$^{1}$
C.~Blake,$^{1,*}$
U.~Andrade,$^{2,3}$
H.~E.~Noriega,$^{4,5}$
J.~Aguilar,$^{6}$
S.~Ahlen,$^{7}$
S.~BenZvi,$^{8}$
D.~Bianchi,$^{9,10}$
D.~Brooks,$^{11}$
T.~Claybaugh,$^{6}$
A.~Cuceu,$^{6}$
A.~de la Macorra,$^{5}$
J.~DeRose,$^{12}$
P.~Doel,$^{11}$
N.~Emas,$^{1}$
S.~Ferraro,$^{6,13}$
J.~E.~Forero-Romero,$^{14,15}$
C.~Garcia-Quintero,$^{16}$
E.~Gaztañaga,$^{17,18,19}$
G.~Gutierrez,$^{20}$
S.~Heydenreich,$^{21}$
K.~Honscheid,$^{22,23,24}$
C.~Howlett,$^{25}$
D.~Huterer,$^{26,3}$
M.~Ishak,$^{27}$
S.~Joudaki,$^{28}$
R.~Joyce,$^{29}$
E.~Jullo,$^{30}$
R.~Kehoe,$^{31}$
D.~Kirkby,$^{32}$
A.~Kremin,$^{6}$
A.~Krolewski,$^{33,34,35}$
O.~Lahav,$^{11}$
A.~Lambert,$^{6}$
C.~Lamman,$^{16}$
M.~Landriau,$^{6}$
J.~U.~Lange,$^{36}$
L.~Le~Guillou,$^{37}$
A.~Leauthaud,$^{21,38}$
M.~Manera,$^{39,40}$
A.~Meisner,$^{29}$
R.~Miquel,$^{41,40}$
S.~Nadathur,$^{18}$
J.~ A.~Newman,$^{42}$
G.~Niz,$^{43,44}$
N.~Palanque-Delabrouille,$^{45,6}$
W.~J.~Percival,$^{33,34,35}$
A.~Porredon,$^{28,46,47,24}$
F.~Prada,$^{48}$
I.~P\'erez-R\`afols,$^{49}$
G.~Rossi,$^{50}$
R.~Ruggeri,$^{51}$
E.~Sanchez,$^{28}$
C.~Saulder,$^{52}$
D.~Schlegel,$^{6}$
A.~Semenaite,$^{1}$
J.~Silber,$^{6}$
D.~Sprayberry,$^{29}$
Z.~Sun,$^{53}$
G.~Tarl\'{e},$^{3}$
B.~A.~Weaver,$^{29}$
P.~Zarrouk,$^{37}$
R.~Zhou,$^{6}$
and H.~Zou$^{54}$ \\
{\it (Affiliations can be found after the references)}
}
\thanks{$^*$E-mail: cblake@swin.edu.au}

\begin{abstract}
The action of gravitational physics across space-time creates observable signatures in the behaviour of light and matter.  We perform combined-probe studies using data from the Baryon Oscillation Spectroscopic Survey (BOSS) and Dark Energy Spectroscopic Instrument survey Data Release 1 (DESI-DR1), in combination with three existing weak lensing surveys, the Kilo-Degree Survey (KiDS), the Dark Energy Survey (DES), and the Hyper Suprime-Cam Survey (HSC), to test and constrain General Relativity (GR) in the context of the standard model of cosmology ($\Lambda$CDM).  We focus on measuring the gravitational estimator statistic, $E_G$, which describes the relative amplitudes of weak gravitational lensing and galaxy velocities induced by a common set of overdensities.  By comparing our amplitude measurements with their predicted scale- and redshift-dependence within the GR+$\Lambda$CDM model, we demonstrate that our results are consistent with the predictions of the \textit{Planck} cosmology.  The redshift span of the DESI dataset allows us to perform these $E_G$ measurements at the highest redshifts achieved to date, $z \approx 1$.
\\[1em]
\end{abstract}

\maketitle

\section{Introduction}

Over the past three decades, the volume and precision of cosmological observations have increased dramatically, leading to a renewed focus on testing Einstein’s General Theory of Relativity (GR). Despite the many successes of GR in describing phenomena from solar-system scales to large-scale structure, the mechanism behind the Universe’s accelerated expansion remains elusive. Since this phenomenon was discovered via Type Ia supernova observations \citep{riess_observational_1998, perlmutter_measurements_1999}, numerous models have emerged to explain cosmic acceleration \citep[for reviews see e.g.,][]{yoo_theoretical_2012, joyce_dark_2016, ishak_testing_2018, ferreira_cosmological_2019}. Broadly, these models fall into two main categories: dark energy and modified gravity.

Dark energy scenarios attribute the acceleration to a component in the matter-energy sector that exerts negative pressure. For instance, in quintessence models or scenarios involving a vacuum energy density, gravitational attraction is effectively counteracted by an additional repulsive force, thereby weakening gravitational binding at large scales \citep[e.g.,][]{frieman_dark_2008, weinberg_observational_2013}. Under dark energy frameworks, GR itself is preserved across all scales and epochs. The simplest such model posits a constant vacuum energy, encapsulated by the cosmological constant $\Lambda$. However, the inferred magnitude of $\Lambda$ diverges significantly from theoretical expectations, posing the well-known “cosmological constant problem” \citep{weinberg_cosmological_1989}.

By contrast, modified gravity theories question the completeness of GR itself, introducing alterations to the gravitational action, additional spacetime dimensions, or modifications to relativistic and non-relativistic conservation laws \citep[for reviews see e.g.,][]{clifton_modified_2012, joyce_dark_2016, ishak_testing_2018, ferreira_cosmological_2019, akrami_modified_2021}. With steadily growing datasets, cosmologists have developed methods that minimize \textit{a priori} model assumptions and allow the data to guide the investigation \citep{heavens_cosmology_2011, astier_observational_2012, kilbinger_cosmology_2015, ferte_testing_2019, li_cosmological_2019}. These phenomenological approaches compare observations directly against a library of theoretical modelling scenarios, thus identifying viable models without committing to a single specific framework in advance.

In this paper, we focus on a perspective that characterizes modifications to gravity through the scalar potentials $\Psi$ and $\Phi$, which describe the perturbed spacetime metric \citep{bardeen_gauge-invariant_1980, bertschinger_one_2011}. In standard (GR+$\Lambda$CDM) models, these potentials are equal ($\Phi = \Psi$), and deviations from GR often manifest as a ratio $\Phi / \Psi \neq 1$ which is also known as “anisotropic stress” \citep{amendola_measuring_2008, amendola_model-independent_2014, cardona_traces_2014, arjona_hints_2020, sobral-blanco_measuring_2021}. Although the potentials themselves are not directly observable, careful parameterizations allow their effects to be probed empirically \citep{koyama_cosmological_2016}.

One particularly interesting empirical measure is the gravitational estimator $E_G$ \citep{zhang_probing_2007, reyes_confirmation_2010}, which consists of an amplitude ratio of weak gravitational lensing observables (sensitive to $\Psi + \Phi$) and galaxy velocities (sensitive to $\Psi$) around the same set of gravitational lenses, to test GR on cosmological scales. Under GR and in the weak-field limit, within certain approximations $E_G$ is scale-independent and its dependence on redshift $z$ is simply predicted by,
\begin{equation}
    E_G(z) = \frac{\Omega_m}{f(z)},
\label{eq:egtheory}
\end{equation}
\citep[see][for a detailed derivation of the approximations inherent in Eq.~\ref{eq:egtheory}]{leonard_testing_2015}, where $\Omega_m$ is the present-day matter density and $f(z)\simeq \Omega_m(z)^{0.55}$ is the linear growth rate of structure \citep{2005PhRvD..72d3529L}, where $\Omega_m(z)$ denotes the matter density parameter as measured by observers at redshift $z$.  The scale-independence and redshift-dependence of $E_G$ hence both provide convenient tests of gravitational physics which may be performed when spectroscopic and weak lensing surveys overlap.

This quantity was first measured by \cite{reyes_confirmation_2010} using Sloan Digital Sky Survey (SDSS) data, yielding $E_G(\bar{z}=0.32)=0.39\pm0.06$, consistent with the GR prediction of $0.4$ at that redshift, where $\bar{z}$ denotes the effective redshift of the spectroscopic galaxy survey used to perform the test.  Subsequent measurements of $E_G$ have employed various lensing and spectroscopic surveys to explore gravitational physics across different redshifts \citep{blake_rcslens_2016, de_la_torre_vimos_2017, alam_testing_2017, amon_kids2dflensgama_2018, singh_probing_2019, jullo_testing_2019, blake_testing_2020}, producing results largely consistent with GR, although with some mild tensions persisting.  Gravitational lensing of the Cosmic Microwave Background has also been used to constrain $E_G$ \citep{pullen_constraining_2016, 2020MNRAS.491...51S, 2021MNRAS.501.1013Z, 2024PhRvD.109h3540W}.

In this study we use existing data from the Baryon Oscillation Spectroscopic Survey Data Release 12 \citep[BOSS-DR12,][]{reid_sdss-iii_2016} and new data from the Dark Energy Spectroscopic Instrument Data Release 1 sample \citep[DESI-DR1,][]{2025arXiv250314745D}, in conjunction with overlapping weak lensing surveys including the Kilo-Degree Survey \citep[KiDS-1000,][]{giblin_kids-1000_2021}, Dark Energy Survey \citep[DES-Y3,][]{gatti_dark_2021} and Hyper-SuprimeCam survey \citep[HSC-Y1 and HSC-Y3,][]{mandelbaum_first-year_2018, li_three-year_2022}, to expand upon previous $E_G$ determinations and provide a comprehensive set of results, including measurements at the highest redshifts yet observed.  We will compare the scale-dependence and redshift-dependence of these measurements to predictions of the standard GR+$\Lambda$CDM model with parameters calibrated by observations by the \textit{Planck} satellite \citep{aghanim_planck_2020}.

The paper is organized as follows: in Sec.~\ref{sec:EGtheory}, we present the theoretical foundations and key ingredients of a direct $E_G$ measurement.  We describe our datasets and simulations in Sec.~\ref{sec:data}, and detail our measurement pipeline in Sec.~\ref{sec:eg_pipeline}. Sec.~\ref{sec:sim} presents simulation tests and their outcomes, while the results of our measurements using the galaxy surveys are reported in Sec.~\ref{sec:egResults}.  In Sec.~\ref{sec:discussion}, we discuss the implications of our findings, place them in the context of existing measurements, and we conclude in Sec.~\ref{sec:conclusion}.

\section{Theoretical foundations}
\label{sec:EGtheory}

In this section we outline the theoretical framework of the $E_G$ estimator, focusing on the main definitions and equations relevant to measuring $E_G$ from galaxy-galaxy lensing observables.  Probing modifications to gravity often requires assumptions about the background geometry and cosmic expansion \citep{skara_tension_2020, arjona_hints_2020}. Here, we adopt a flat-$\Lambda$CDM framework with matter density $\Omega_m$ and gravitational physics governed by GR. Significant observed deviations from model-based expectations can then indicate a breakdown of one or more of these underlying assumptions.

\subsection{Tangential shear \texorpdfstring{$\gamma_t$}{gamma\_t}}
\label{subsec:egGamma_t}

The average tangential shear, $\gamma_t(\theta)$, quantifies the mean tangential distortion of background (source) galaxy shapes caused by foreground (lens) galaxies. At an angular separation $\theta$, it may be defined as the expectation value of the product between the overdensity in the lens galaxy field, $\delta_g(\mathbf{x})$, and the tangential shear observed at a relative position $\mathbf{x}+\theta$,
\begin{equation}
\label{eq:tanshear1}
    \gamma_t(\theta) = \langle \,\delta_g(\mathbf{x})\, \gamma_t(\mathbf{x}+\theta)\rangle .
\end{equation}
In practice this quantity is computed by averaging the projected source shape over all source-lens pairs within a chosen separation bin.

From a cosmological perspective, the tangential shear can also be written as a Hankel transform of the galaxy-convergence cross-power spectrum, $C_{g\kappa}(\ell)$,
\begin{equation}
\label{eq:gammat_hankel}
    \gamma_t(\theta) = \frac{1}{2\pi}\int_{0}^{\infty} \,\ell \,C_{g\kappa}(\ell)\, J_2(\ell \theta)\,\mathrm{d}\ell ,
\end{equation}
where $J_2$ is the second-order Bessel function of the first kind. The cross-power spectrum itself can be written (under a spatially flat assumption) as,
\begin{equation}
\begin{split}
\label{eq:cgkappa}
    C_{g\kappa}(\ell) &= \frac{3\,\Omega_m\,H_0^2}{2\,c^2} \times \\
    &\int_{0}^{\infty} \frac{(1+z)}{\chi(z)}\,P_{gm}\left(\frac{\ell+\frac{1}{2}}{\chi(z)},z\right)\, p_l(z)\, W(z)\,\mathrm{d}z,
\end{split}
\end{equation}
in which the lensing efficiency $W(z)$ is given by,
\begin{equation}
\label{eq:lenseff}
    W(z) = \int_{z}^{\infty}p_s(z')\, \left[ \frac{\chi(z')-\chi(z)}{\chi(z')} \right] \,\mathrm{d}z'.
\end{equation}
Here, $\chi(z)$ is the comoving distance, $\Omega_m$ is the matter density parameter, $H_0$ is the Hubble constant, $c$ is the speed of light, and $P_{gm}(k,z)$ denotes the galaxy-matter cross-power spectrum. The functions $p_s(z)$ and $p_l(z)$ represent the normalised redshift distributions of the source and lens samples, respectively.

\subsection{Differential surface density \texorpdfstring{$\Delta \Sigma(R)$}{Delta Sigma(R)}}
\label{subsec:Dsigma}

The differential surface mass density, $\Delta \Sigma(R)$, quantifies the projected mass distribution around a lens in comoving coordinates. It is directly related to the tangential shear, $\gamma_t(\theta)$, through the comoving critical surface mass density, $\Sigma_c$.  For a single lens-source pair at redshifts $z_l$ and $z_s$, the tangential shear at tranverse separation $R$, corresponding to an angular separation $\theta = R/\chi(z_l)$, can be written as,
\begin{equation}
\label{eq:tanshear2}
    \gamma_t(\theta) = \frac{\Delta \Sigma(R, z_l)}{\Sigma_c(z_l, z_s)} ,
\end{equation}
with the comoving critical surface density defined by,
\begin{equation}
\label{eq:sigmac}
    \Sigma_c(z, z') = \frac{c^2}{4 \pi G} \left[\frac{\chi(z')}{(\chi(z')-\chi(z))\,\chi(z)\,(1+z)}\right] ,
\end{equation}
where $G$ is the gravitational constant, and $z' > z$.

The differential surface mass density itself is given by,
\begin{equation}
\label{eq:excessDSigma}
    \Delta \Sigma(R) = \overline{\Sigma}(< R) - \Sigma(R) ,
\end{equation}
where $\overline{\Sigma}(< R)$ is the mean projected mass density within radius $R$. In terms of the galaxy-matter cross-correlation function, $\xi_{gm}(r)$, we may write,
\begin{equation}
\label{eq:DeltaSigma_xigm}
    \Delta \Sigma(R) = \overline{\rho}_m \int_{-\infty}^{\infty} \xi_{gm}(\sqrt{R^2 + \Pi^2})\,\mathrm{d}\Pi,
\end{equation}
where $\overline{\rho}_m = \rho_c \,\Omega_m$ is the mean comoving matter density, in terms of the critical density $\rho_c = 3 H_0^2 / 8 \pi G$, and $\Pi$ is the LoS separation.

In a linear bias framework, $\xi_{gm}(r)$ can be obtained from the non-linear matter power spectrum $P_{mm}(k,z)$, with a scale-independent galaxy bias factor $b(z)$,
\begin{equation}
\label{eq:xi_gm}
    \xi_{gm}(r, z) = \frac{b(z)}{2\pi^2} \int_{0}^{\infty} k^2\,P_{mm}(k,z)\,\frac{\sin(kr)}{kr}\,\mathrm{d}k.
\end{equation}
where we generate $P_{mm}(k,z)$ using the \textsc{camb} software \citep{2000ApJ...538..473L} including the \textsc{halofit} corrections \citep{2003MNRAS.341.1311S} as re-calibrated by \cite{2012ApJ...761..152T}.  The fiducial cosmological parameter sets we use for analysis of the mock and real datasets are collected in Table~\ref{tab:parameters}.  By relating Eq.~\ref{eq:gammat_hankel} and Eq.~\ref{eq:tanshear2} from Sec.~\ref{subsec:egGamma_t} we obtain,
\begin{equation}
\label{eq:dstogt}
    \gamma_t(\theta) = \int_{0}^{\infty} p_l(z)\,\Delta\Sigma(R,z) \, \mathrm{d}z \\ \int_{z}^{\infty} \frac{p_s(z')}{\Sigma_c(z,z')} \, \mathrm{d}z' .
\end{equation}
In the limit of narrow redshift distributions, $p_s(z) = \delta_D(z - z_s)$ and $p_l(z)=\delta_D(z-z_l)$, the above expression reduces to Eq.~\ref{eq:tanshear2}. For a narrow lens distribution and averaging over the source redshift distribution we find,
\begin{equation}
\label{eq:gamma_t_DelSig}
    \Delta \Sigma(R, z_l) \simeq \langle \gamma_t(\theta)\rangle \left[\int_{z_l}^{\infty} \frac{p_s(z')}{\Sigma_c(z_l,z')} \, \mathrm{d}z' \right]^{-1} .
\end{equation}

\begin{table}
	\centering
	\caption{Fiducial parameter values assumed for our analyses of the Buzzard mocks and BOSS/DESI dataset.}
	\label{tab:parameters}
	\begin{tabular}{c|c|c}
		\hline 
		\textbf{Parameter} & \textbf{Buzzard} & \textbf{BOSS/DESI} \\ 
        & \textbf{fiducial} & \textbf{fiducial} \\
		\hline
            Matter density $\Omega_m$ & $0.286$ & $0.3153$ \\
            Baryon density $\Omega_b$ & $0.046$ & $0.0493$ \\
            Hubble parameter $h$ & $0.7$ & $0.6736$ \\
            Normalisation $\sigma_8$ & $0.82$ & $0.8114$ \\
            Spectral index $n_s$ & $0.96$ & $0.9649$ \\
        \hline
            Intrinsic alignment amplitude $A_{\rm IA}$ & 0 & 0.5 \\
            Lens magnification $\alpha_l$ & 0 & 2 \\
        \hline
    \end{tabular}
\end{table}

\subsection{Projected clustering \texorpdfstring{$w_p(R)$}{w\_p(R)} and redshift-space distortions}
\label{subsec:egClustering}

Galaxy velocities are incorporated in the $E_G$ statistic by using the observed amplitude of redshift-space distortions (RSD), normalised by the projected radial profile of galaxy clustering.  When measuring 3D clustering in redshift space, galaxy positions are influenced by peculiar velocities along the line-of-sight (LoS). This effect modifies the observed galaxy overdensity in redshift space, $\delta_g^s(\mathbf{x})$, whose Fourier transform can be written as,
\begin{equation}
    \tilde{\delta}_g^s(k,\mu) = \tilde{\delta}_g(k) - \mu^2\,\tilde{\theta}(k) ,
\end{equation}
where $\mu$ is the cosine of the angle between the wavevector $\mathbf{k}$ and the LoS, and $\tilde{\theta}(k)$ is the Fourier transform of the velocity field divergence in units of the Hubble comoving velocity, $\theta(\mathbf{x}) = \nabla \cdot \mathbf{v}(\mathbf{x})/Ha$. Under linear perturbation theory,
\begin{equation}
    \tilde{\theta}(k) = -\,f\,\tilde{\delta}_m(k) ,
\end{equation}
where $f \equiv \mathrm{d}\ln D/\mathrm{d}\ln a \,$ is the linear growth rate of structure, and $D(a)$ is the linear growth factor for the matter overdensity $\delta(a)$.

Assuming a linear galaxy bias $b$ such that $\delta_g(\mathbf{x}) = b\,\delta_m(\mathbf{x})$, the redshift-space galaxy power spectrum in the linear regime becomes,
\begin{equation}
\label{eq:P_gg}
    P_{gg}^s(k,\mu) = b^2(1+\beta\,\mu^2)^2\,P_{mm}(k) ,
\end{equation}
where $\beta = f/b$ governs the strength of redshift-space distortions, which hence may be inferred by the observed dependence of the clustering amplitude on the angle to the LoS.

The 3D galaxy auto-correlation function, $\xi_{gg}(\mathbf{r})$, can be recovered by the inverse Fourier transform of $P_{gg}(\mathbf{k})$,
\begin{equation}
    \xi_{gg}(\mathbf{r}) = \frac{1}{(2\pi)^3} \int P_{gg}(\mathbf{k})\, e^{-i \mathbf{k}\cdot\mathbf{r}} \, \mathrm{d}^3\mathbf{k} .
\end{equation}
When estimating $E_G$, we separate the RSD amplitude from the projected correlation function $w_p(R)$ which traces the density profile around galaxies independently of RSD,
\begin{equation}
    w_p(R) = \int_{-\infty}^{\infty} \xi_{gg}(R,\Pi) \, \mathrm{d}\Pi .
\end{equation}
Hence, the observed quantities used in an $E_G$ determination are the differential surface density $\Delta\Sigma(R)$, the projected galaxy correlation function $w_p(R)$, and the redshift-space distortion parameter $\beta$.

\subsection{Suppressing non-linear effects}
\label{subsec:linearity}

On small scales, non-linearities in galaxy bias and the galaxy-mass correlation become significant, complicating the theoretical modelling of galaxy-galaxy lensing and clustering. To mitigate these difficulties, \cite{mandelbaum_precision_2010} and \cite{2010PhRvD..81f3531B} introduced annular statistics that subtract contributions from scales below a chosen cut-off $R_0$, where ``one-halo'' terms dominate. Specifically, the annular differential surface density $\Upsilon_{gm}(R)$ is defined as,
\begin{equation}
\label{eq:Ygm_theory}
\begin{split}
    \Upsilon_{gm}(R,R_0) &= \frac{2}{R^2} \int_{R_0}^{R} R'\,\Sigma(R') \,\mathrm{d}R' \\ &-\Sigma(R) + \frac{R_0^2}{R^2}\,\Sigma(R_0) ,
\end{split}
\end{equation}
where $R_0$ is an independently chosen inner radius corresponding approximately to the one-halo scale. By construction, $\Upsilon_{gm}(R_0)=0$, ensuring that highly non-linear contributions below $R_0$ are suppressed.

An analogous approach is used to define the annular projected galaxy clustering density, $\Upsilon_{gg}$,
\begin{equation}
\label{eq:Ygg_theory}
\begin{split}
    &\Upsilon_{gg}(R,R_0) = \rho_c\, \times \\
    &\left[\, \frac{2}{R^2} \int_{R_0}^{R} R'\, w_p(R') \,\mathrm{d}R' - w_p(R)
    + \frac{R_0^2}{R^2}\,w_p(R_0) \right] .
\end{split}
\end{equation}
This annular statistic similarly reduces the impact of non-linear, small-scale contributions to $w_p(R)$, enhancing the reliability of comparisons to linear-theory predictions.  To connect these annular statistics to potential deviations from GR, we next examine how modifications of gravity can manifest through the gravitational slip parameter and the $E_G$ estimator.

\subsection{The relationship between gravitational slip \texorpdfstring{$\eta$}{eta} and the \texorpdfstring{$E_G$}{E\_G} statistic}
\label{subsec:standardEG}

A convenient way to parameterise potential departures from GR is via the gravitational slip, $\eta$, which quantifies the difference between the lensing potential $\Phi$ and the Newtonian potential $\Psi$. Defining $\eta$ by,
\begin{equation}
\label{eq:eta}
    \frac{\Psi + \Phi}{\Psi} = 1 + \eta\, ,
\end{equation}
we obtain $\eta = \Phi/\Psi \ne 1$ in the presence of anisotropic stress. In standard GR, typically $\Phi=\Psi$, so $\eta=1$ \footnote{We note that anisotropic stress could be sourced by relativistic species even within GR \citep{2010PhRvD..81j4023P}.}. Any $\eta \neq 1$ indicates new physics \citep{amendola_measuring_2008, amendola_model-independent_2014, cardona_traces_2014, joyce_dark_2016, sobral-blanco_measuring_2021}.

The $E_G$ statistic provides an observationally accessible means of searching for gravitational slip, connecting lensing observables to peculiar velocity fields.  \cite{zhang_probing_2007} proposed a theoretical definition,
\begin{equation}
\label{eq:theoryEG}
    E_G \equiv \left[ \frac{ \tilde{\nabla}^2 (\Psi + \Phi)} {3\,H_0^2\,a^{-1}\,\tilde{\theta}} \right]_{k=\ell/\bar{\chi},\,\bar{z}} ,
\end{equation}
where $\ell$ is a transverse wave number and $\bar{\chi}$ is the mean comoving distance at redshift $\bar{z}$. In the weak-field, sub-horizon limit of GR we have,
\begin{equation}
    \nabla^2 \Phi_{GR} = \nabla^2 \Psi_{GR} = 4\pi G a^2 \rho \delta = \tfrac{3}{2} H_0^2 \Omega_m a^{-1} \delta ,
\end{equation}
which yields the standard GR prediction for $E_G$ given in Eq.~\ref{eq:egtheory}, where caveats and approximations are noted by \cite{leonard_testing_2015}.

Observationally, \cite{reyes_confirmation_2010} introduced a practical estimator for $E_G$ in terms of the annular statistics and redshift-space distortion parameter, 
\begin{equation}
\label{eq:eg_numerical}
    E_G(R,z) = \frac{1}{\beta(z)} \frac{\Upsilon_{gm}(R,z)}{\Upsilon_{gg}(R,z)} .
\end{equation}
The linear-theory dependencies of $\beta$, $\Upsilon_{gm}(R)$ and $\Upsilon_{gg}(R)$ on galaxy bias $b$ and the amplitude of matter clustering $\sigma_8$ cancel out in this ratio (given that $\beta \propto 1/b$, $\Upsilon_{gm}(R) \propto b \sigma_8^2$ and $\Upsilon_{gg}(R) \propto b^2 \sigma_8^2$), yielding a measurement sensitive primarily to departures from GR \citep{zhang_probing_2007, simpson_cfhtlens_2013, skara_tension_2020, arjona_hints_2020}, although we note that cosmological parameters such as $\Omega_m$ do not cancel and can hence affect the ratio \citep{amon_kids2dflensgama_2018}.  Nevertheless, this formalism makes $E_G$ a powerful tool to detect either a scale dependence or a redshift-dependent amplitude shift relative to GR, potentially pointing toward new physics if significant discrepancies arise.

\subsection{Modelling other astrophysical effects}

The measured galaxy-galaxy lensing signal is also influenced by astrophysical effects, which we wish to model and subtract in order to isolate the contribution of weak gravitational lensing.  We focus on two such effects in particular: the intrinsic alignment (IA) of galaxies, whereby their shapes align coherently due to local tidal fields rather than due to lensing \citep[for a recent review see,][]{2024OJAp....7E..14L}, and lens magnification, which describes the fluctuation in observed lens density caused by lens selection above a fixed magnitude limit \citep[see e.g.,][]{2020A&A...638A..96U}.

We can estimate the contribution of intrinsic alignments to the galaxy-galaxy lensing signal as \citep[e.g.,][]{2023PhRvD.108l3521S},
\begin{equation}
\label{eq:dsia}
    \Delta \Sigma_{\rm IA}(R) = \frac{1}{2\pi} \int_0^\infty k \, J_2(kR) \, P_{\rm IA}(k) \, dk ,
\end{equation}
where the associated power spectrum of the intrinsic alignment contribution is given by,
\begin{equation}
    P_{\rm IA}(k) = b \int_0^\infty p_l(z) f_{\rm IA}(z) \left[ \overline{\Sigma_c^{-1}} \right]^{-1} P_{mm}(k,z) \, dz ,
\label{eq:pkia}
\end{equation}
where $\overline{\Sigma_c^{-1}}$ is the average inverse critical density of the source sample relative to lens redshift $z$,
\begin{equation}
    \overline{\Sigma_c^{-1}}(z) = \int_z^\infty p_s(z_s) \, \Sigma_c^{-1}(z,z_s) \, dz_s .
\end{equation}
We evaluated the intrinsic alignment kernel $f_{\rm IA}(z)$ in Eq.~\ref{eq:pkia} using the non-linear alignment (NLA) model \citep{2007NJPh....9..444B},
\begin{equation}
    f_{\rm IA}(z) = - A_{\rm IA} \, \left( \frac{1+z}{1+z_{\rm piv}} \right)^\eta \, \frac{C_1 \, \rho_c \, \Omega_m}{D(z)} \, \frac{p_s(z)}{d\chi/dz} ,
\end{equation}
where $A_{\rm IA}$, $z_{\rm piv}$, $\eta$ and $C_1$ are model parameters.  For our estimate, we assume standard choices with no evolution ($\eta = 0$), $C_1 \rho_c = 0.0134$, and $A_{\rm IA} = 0.5$ as a representative value of the intrinsic alignment amplitude \citep[following][]{2024OJAp....7E..57L}.

We estimated the contribution of lens magnification to the galaxy-galaxy lensing signal, $\Delta \Sigma_{\rm mag}(R)$, by evaluating the equivalent expression to Eq.~\ref{eq:dsia} using a lens magnification power spectrum \citep{2023PhRvD.108l3521S},
\begin{equation}
\begin{split}
    P_{\rm mag}(k) &= 2 \left( \alpha_l - 1 \right) \left( \frac{3 \Omega_m H_0^2}{2 c^2} \right)^2 \times \\ &\int_0^\infty (1+z)^2 \, \chi^2 \, f_{\rm mag}(z) \, P_{mm}(k,z) \, \frac{d\chi}{dz} \, dz ,
\end{split}
\end{equation}
where $\alpha_l$ is the lens magnification parameter, and the lens magnification kernel $f_{\rm mag}(z)$ is given by,
\begin{equation}
\begin{split}
    f_{\rm mag}(z) &= \int_z^\infty dz_l \, p_l(z_l) \left( \frac{\chi_l - \chi(z)}{\chi_l} \right) \\ &\int_{z_l}^\infty dz_s \, p_s(z_s) \left( \frac{\chi_s - \chi(z)}{\chi_s} \right) \, \Sigma_c(z_l, z_s) ,
\end{split}
\end{equation}
where we assume $\alpha_l = 2$ as a value representative of our data \citep[see,][]{2021MNRAS.504.1452V, 2025arXiv250621677H}.

We always evaluate and include these two astrophysical model contributions to the galaxy-galaxy lensing signal for each pair of source and lens tomographic bins, which we hence write as,
\begin{equation}
\Delta\Sigma(R) = \Delta\Sigma_{\rm GGL}(R) + \Delta\Sigma_{\rm IA}(R) + \Delta\Sigma_{\rm mag}(R) ,
\end{equation}
although the contributions of the intrinsic alignment and magnification terms are negligible for the scales and source-lens pairs included in our analysis.  Switching off these corrections entirely, the average absolute change in the $E_G(z)$ measurements relative to the errors is $0.1 \sigma$, compared to our fiducial correction; marginalising over a range of values for the intrinsic alignment or magnification parameters within currently determined limits would not significantly change our $E_G$ measurements or errors.  We refer the reader to \cite{2024OJAp....7E..57L} for full simulation studies of these effects in the context of DESI.

\section{Data}
\label{sec:data}

\begin{figure*}
    \centering
    \includegraphics[width=2\columnwidth]{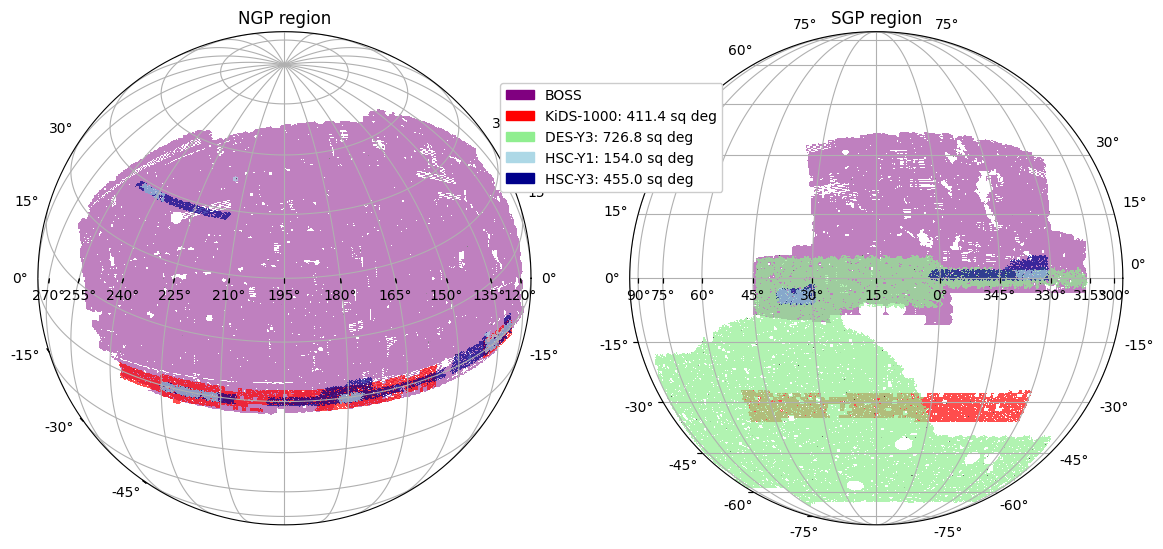}
    \includegraphics[width=2\columnwidth]{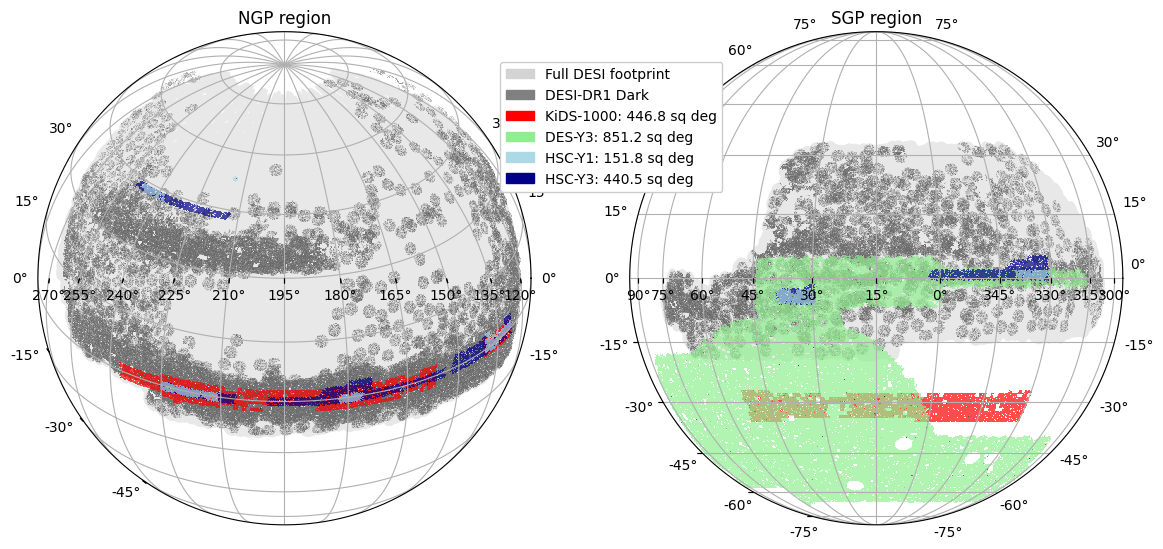}
    \caption{The sky coverage of BOSS-DR12 (top) and DESI-DR1 (bottom), showing the overlapping footprints for KiDS-1000 (red), DES-Y3 (green), HSC-Y1 (light blue) and HSC-Y3 (dark blue), projected onto the celestial sphere using equatorial coordinates (right ascension and declination).  Two hemispheres are shown in each case, centred on the North Galactic Pole (NGP) and South Galactic Pole (SGP) regions.  The full DESI footprint is shown in the bottom panel as the light grey region.}
    \label{fig:footprints}
\end{figure*}

\begin{table*}
  \centering
  \caption{The overlap areas and the number of lenses used from the BOSS-DR12, DESI-DR1-BGS and DESI-DR1-LRG datasets in combination with the KiDS-1000, DES-Y3, HSC-Y1 and HSC-Y3 weak lensing surveys.  The HSC-Y3 dataset only became available after we had completed the analysis using BOSS, so this combination is not included in our study.}
    \begin{tabular}{p{0.6cm}|p{2cm}|c|c|c|c|}
    \cline{3-6}
    \multicolumn{1}{c}{}&\multicolumn{1}{c|}{} & \centering\arraybackslash\textbf{KiDS-1000}  & \centering\arraybackslash\textbf{DES-Y3} & \centering\arraybackslash\textbf{HSC-Y1} &
    \centering\arraybackslash\textbf{HSC-Y3}\\
    \cline{2-6}   
    \multirow{5.5}{*}{\rotatebox[origin=c]{90}{}} & \centering\textbf{BOSS-DR12} & $411.4$ deg$^2$  & $726.8$ deg$^2$ & $154.0$ deg$^2$ & -- \\
     & \centering\text{(LRG)} & $47{,}332$ & $104{,}008$ & $24{,}667$ & --\\    
     \cline{2-6}  
     & \centering\textbf{DESI-DR1} & $448.2$ deg$^2$  & $716.8$ deg$^2$ & $142.1$ deg$^2$ & $440.6$ deg$^2$ \\
     & \centering\text{(BGS)} & $22{,}073$ & $23{,}465$ & $6{,}751$ & $18{,}227$ \\
    \cline{2-6}  
     & \centering\textbf{DESI-DR1} & $446.8$ deg$^2$  & $851.2$ deg$^2$ & $151.8$ deg$^2$ & $440.5$ deg$^2$ \\ 
     & \centering\text{(LRG)} & $205{,}139$ & $181{,}298$ & $59{,}239$ & $168{,}552$ \\
    \cline{2-6} 
    \end{tabular}
  \label{tab:SurveyCoverage}
\end{table*}

\subsection{DESI}
\label{subsec:desiDataIntro}

As the primary large-scale structure dataset in this study, we use the Dark Energy Spectroscopic Instrument Data Release 1 \citep[DESI-DR1,][]{2025arXiv250314745D}.  Over its five-year program, DESI aims to collect over 40 million spectra of galaxies and quasars across $14{,}000$\,deg$^2$, to measure the expansion history of the Universe and the growth of large-scale structure \citep{2013arXiv1308.0847L, desi_collaboration_desi_2016, 2016arXiv161100037D, 2024AJ....167...62D}.  The overall DESI observing strategy is summarised by \cite{2023AJ....166..259S}.  The survey obtains spectra for four principal target classes: the Bright Galaxy Survey (BGS), Luminous Red Galaxy Survey (LRG), Emission Line Galaxy Survey (ELG), and the Quasar Survey (QSO), which are photometrically-selected from the DESI Legacy optical imaging surveys \citep{2019AJ....157..168D}.  The instrument design is summarised by \cite{2022AJ....164..207D}: targets are assigned to 5000 optical fibres in the telescope focal plane \citep{2024AJ....168..245P} using a robotic positioner \citep{2023AJ....165....9S}, and the data are processed by the DESI spectroscopic pipeline \citep{2023AJ....165..144G}.  

In our analysis we use the clustering data and random catalogues from the Bright Galaxy Survey \citep{BGS1, BGS2} and Luminous Red Galaxy survey \citep{LRG2}, which form the most efficient gravitational lenses for background sources owing to their location at $z < 1$ and in the most massive dark matter halos.  We divide these samples by redshift to form six DESI lens samples.  Bright Galaxies occupy $0.1 < z < 0.4$, which we divide into three redshift bins of width $\Delta z = 0.1$, and LRGs occupy $0.4 < z < 1.1$, divided into three redshift bins $[0.4,\,0.6]$, $[0.6,\,0.8]$ and $[0.8,\,1.1]$.  The overlap of the DESI data and the KiDS-1000, DES-Y3, HSC-Y1 and HSC-Y3 weak lensing surveys is displayed in Fig.~\ref{fig:footprints} and detailed in Table~\ref{tab:SurveyCoverage}.  We note that in our analysis we apply the same absolute magnitude cut $M_r < -21.5$ to the BGS sample as used to define the DESI Key Project catalogues \citep{2025JCAP...07..017A}, given that we will utilise the same redshift-space distortion measurements as published for this sample \citep{2025JCAP...09..008A}.  In this respect our BGS lens samples differ from those analysed by \cite{2025arXiv250621677H}, who create fainter samples in each redshift bin.

\subsection{BOSS}
\label{subsec:bossDataIntro}

We compare our measurements using DESI with its principal predecessor, the Baryon Oscillation Spectroscopic Survey \citep[BOSS,][]{2013AJ....145...10D}, which was part of the Sloan Digital Sky Survey III \citep[SDSS-III,][]{2011AJ....142...72E}.  Its twelfth data release (DR12) contains about 1.5 million galaxy and quasar spectra over $\sim10{,}000$\,deg$^2$ \citep{reid_sdss-iii_2016}.  BOSS primarily targeted LRGs based on optical colour selections from SDSS imaging. Our analysis focuses on BOSS LRGs in the redshift interval $0.2 < z < 0.7$.  For $z<0.2$ and $z>0.7$, the LRG number density decreases significantly.  We divide this redshift range into five equal-width lens bins with central redshifts $\bar{z} = [0.25,\,0.35,\,0.45,\,0.55,\,0.65]$.  The overlaps between the BOSS footprint and weak lensing surveys are depicted in Fig.~\ref{fig:footprints} and Table~\ref{tab:SurveyCoverage}.

\subsection{KiDS}
\label{subsec:kidsDataIntro}

We jointly analyse these two galaxy redshift surveys with three weak lensing surveys.  First, we utilise the Kilo-Degree Survey \citep[KiDS-1000,][]{kuijken_fourth_2019}. Observations for KiDS use the OmegaCAM camera on the 2.6\,m VLT Survey Telescope (VST) at ESO's Paranal Observatory in Chile.  KiDS employs four broadband photometric filters ($u, g, r, i$), further extended by adding near-infrared bands ($Y, J, H, K$) from the VISTA Kilo-Degree Infrared Galaxy Survey (VIKING). This allows the KiDS-1000 source catalogue to be divided into five photometric redshift bins with bin limits $z_p = [0.1, 0.3, 0.5, 0.7, 0.9, 1.2]$ \citep{giblin_kids-1000_2021}.  Shape measurements are performed using the \textsc{lensfit} algorithm \citep{miller_bayesian_2007}.  The KiDS-1000 data release comprises $1{,}006$\,deg$^2$ of imaging with around $21$ million galaxy shapes, with effective number density $n_{\rm eff}=6.17$\,arcmin$^{-2}$ and shape noise $\sigma_e=0.265$ \citep{giblin_kids-1000_2021}.  We used the original KiDS-1000 source redshift distribution calibrations in each tomographic bin \citep{2021A&A...647A.124H}, noting that these were recently revised by \cite{2025arXiv250319440W}.

\subsection{DES}
\label{subsec:desDataIntro}

The second weak lensing survey we consider in this study is the Dark Energy Survey Year 3 sample \citep[DES-Y3,][]{gatti_dark_2021}.  DES observations utilize the 570-megapixel Dark Energy Camera (DECam) on the 4\,m Blanco telescope at Cerro Tololo Inter-American Observatory in Chile. With a $3$\,deg$^2$ field of view, DES covers approximately $5{,}000$\,deg$^2$.  Photometry is obtained in five broadband filters ($g, r, i, z, Y$), which collectively enable photometric redshift estimates of source galaxies in the range $z_p < 2$.  For the weak lensing component, DES divides this interval into four tomographic bins with bin limits $z_p = [0, 0.36, 0.63, 0.87, 2.0]$ \citep{gatti_dark_2021}.  Galaxy shape measurements in DES-Y3 are performed using \textsc{metacalibration} \citep{huff_metacalibration_2017, sheldon_practical_2017}, where small artificial shears are applied to each image to determine the shear response.   The DES-Y3 cosmic shear catalogue comprises $\sim 100$ million sources across $4{,}143$\,deg$^2$, which represents a number density $n_{\rm eff}=5.0$\,arcmin$^{-2}$ and shape noise $\sigma_e=0.26$ \citep{gatti_dark_2021}.

\subsection{HSC}
\label{subsec:hscDataIntro}

The final lensing dataset we utilise is the Hyper Suprime-Cam (HSC) survey \citep{mandelbaum_first-year_2018, aihara_third_2022}.  HSC observations are carried out using the 1.77\,deg$^2$ HSC camera at the 8.2\,m Subaru telescope on Maunakea, Hawaii.  The HSC Subaru Strategic Program (SSP) will ultimately cover about $1{,}100$\,deg$^2$ in its Wide tier, using five broadband filters ($g,r,i,z,Y$). Shape measurements are derived predominantly from $i$-band imaging under good seeing conditions using the \textsc{reGauss} PSF correction \citep{hirata_shear_2003, li_three-year_2022} implemented in \textsc{GalSim} \citep{rowe_galsim_2015}, and the dataset is divided into four tomographic bins across photometric redshift range $0.3 < z_p < 1.5$ with bin limits $z_p = [0.3, 0.6, 0.9, 1.2, 1.5]$.

The first-year HSC shear catalogue (HSC-Y1) spans 136.9\,deg$^2$ with an effective galaxy number density of $n_{\rm eff}=21.8$\,arcmin$^{-2}$ and shape noise $\sigma_e \approx 0.28$ \citep{mandelbaum_first-year_2018}.  The more recent HSC-Y3 dataset covers an expanded area of 416\,deg$^2$ with an effective number density of $n_{\rm eff}\approx19.9$\,arcmin$^{-2}$ \citep{li_three-year_2022}.  We note that the HSC-Y3 dataset only became available after we had completed our analysis using BOSS, so is only included in our set of DESI cross-correlation measurements.

\subsection{Buzzard DESI mocks}
\label{subsec:buzzard}

In order to test the analysis pipeline used in this study, we utilised mock catalogues generated from the \textit{Buzzard} simulation suite \citep{derose_buzzard_2019}, which allows us to create representative datasets for both spectroscopic and weak lensing surveys.  Specifically, our mocks include DESI-like BGS and LRG samples and simulated weak lensing catalogues resembling KiDS-1000, DES-Y3 and HSC-Y1.  Below, we summarize the main features of these simulations; further details can be found in \cite{derose_buzzard_2019}, \cite{2024OJAp....7E..57L} and \cite{2025OJAp....8E..24B}.

The Buzzard mocks provide hemisphere-scale light cones drawn from an $N$-body simulation in a fiducial flat GR+$\Lambda$CDM cosmology with parameter values collected in Table~\ref{tab:parameters}.  A halo occupation distribution (HOD) model is employed to populate halos with galaxies in a manner consistent with the projected clustering of early DESI observations \citep{2024AJ....167...62D}.  Likewise, weak lensing catalogues are constructed to reproduce the source redshift distributions, galaxy magnitudes, shape noise, shear calibration biases, and effective number densities of the KiDS-1000, DES-Y3 and HSC-Y1 surveys \citep{2024OJAp....7E..57L}.  We note that the redshift calibration data to construct corresponding mock catalogues for HSC-Y3 was not publicly available at the time of writing, so we did not generate Buzzard lensing mocks corresponding to HSC-Y3.  We use a single Buzzard light cone for the tests carried out in this paper.

The mock data in the light cone are divided into independent regions that are representative of the overlap between DESI and the three weak lensing surveys \citep{2025OJAp....8E..24B}.  In particular, we created $(20, 12, 60)$ regions for the (KiDS-1000, DES-Y3, HSC-Y1) mock catalogues, which each have areas $(483, 806, 161)$ deg$^2$, respectively.  Averaging results over these independent regions allows a more precise test of the performance of our analysis pipeline.  The DESI-like mocks are split into the same six BGS and LRG redshift bins adopted in the DESI analysis: $[0.1,0.2]$, $[0.2,0.3]$ and $[0.3,0.4]$ for BGS, and $[0.4,0.6]$, $[0.6,0.8]$ and $[0.8,1.1]$ for LRGs.

We note that on large scales, the clustering within the Buzzard simulation is impacted by the domain decomposition used when constructing the mocks, as discussed by \cite{derose_buzzard_2019} and \cite{2022ApJ...931..145W}.  Owing to this issue, we cannot use the simulation to perform a redshift-space distortion analysis.  Hence, we will assume a fiducial estimate of the corresponding redshift-space distortion parameters, as discussed below.

\section{Measurements}
\label{sec:eg_pipeline}

\subsection{Source-lens selections}
\label{subsec:sourceLensCuts}

We restrict our analysis to combinations of source and lens tomographic samples for which the source redshift distribution is not located at lower redshifts than the corresponding lens redshift interval, ensuring galaxy-galaxy lensing signal is available. For a lens bin defined over spectroscopic redshift $z_{s,\min}<z_s<z_{s,\max}$, we exclude any source photometric redshift bin $z_{p,\min}<z_p<z_{p,\max}$ if $z_{p,\max} < z_{s,\min}$. In practice, this cut primarily affects higher lens redshift bins.

\subsection{\texorpdfstring{$\Delta \Sigma(R)$}{Delta Sigma (R)} measurements}
\label{subsec:DeltaSigmaMeasurements}

To estimate the differential surface density, $\Delta\Sigma(R)$, for each combination of lens and source tomographic samples, we used the open-source Python package \texttt{dsigma}\footnote{\url{https://dsigma.readthedocs.io/en/latest/}} \citep{2022ascl.soft04006L}.  For the DESI and Buzzard analyses, we compute $\Delta \Sigma(R)$ in 15 logarithmic radial bins spanning $0.08 < R < 80\,h^{-1}\,\mathrm{Mpc}$, and for the BOSS analysis we adopt 10 logarithmic bins between $0.5 < R < 50\,h^{-1}\,\mathrm{Mpc}$ to allow for comparisons with previous work \citep{blake_testing_2020}.  We apply the same small-scale non-linear suppression scale $R_0$ in all cases (as described below).  For each lens $l$ and source $s$ in a given scale bin, the tangential shear of the source ellipticity $e_t$ relative to the lens is converted into an excess surface density,
\begin{equation}
    \Delta \Sigma(R) = \frac{\sum_{ls}\, w_{ls}\,\Sigma_c(z_l,z_s)\, e_t(\theta)}{\sum_{ls}\, w_{ls}} ,
\label{eq:dsigest}
\end{equation}
where $\theta = R / \chi(z_l)$ maps the projected separation $R$ to an angular scale via the lens comoving distance $\chi(z_l)$. The weight assigned to each lens-source pair is,
\begin{equation}
    w_{ls} = \frac{w_s}{\Sigma_c^2(z_l,z_s)} ,
\end{equation}
where $w_s$ is the source galaxy weight derived from shape measurements in the weak lensing catalogues, and $\Sigma_c$ is the comoving critical surface density for the lens-source geometry defined in Eq.~\ref{eq:sigmac}.  Our full implementation of the galaxy-galaxy lensing measurements is described by \cite{2025arXiv250621677H}, in which we apply corrections to account for the multiplicative shear bias calibrated for each weak lensing survey using image simulations (and introduced to the mocks), and for the dilution caused by photometric redshift errors \citep[sometimes known as the $f_{\rm bias}$ factor,][]{2012MNRAS.420.3240N}.  For each source sample, we also subtract the contribution of Eq.~\ref{eq:dsigest} around random lens catalogues with the same distribution as the data.

\subsection{Analytical covariance}
\label{subsec:covMats}

We modelled the correlated errors in the $\Delta\Sigma(R)$ measurements between scales, lens and source samples using an analytical Gaussian covariance as described and validated by \cite{yuan_redshift_2024}.  The analytical covariance calculation includes the sample variance, noise and mixed contributions, in addition to a correction for survey footprint (boundary) effects calibrated using lognormal realisations.  We used the redshift distributions measured for each lens sample or specified for the source samples by the weak lensing collaborations, and an estimate of the linear galaxy bias factor of each lens sample derived from the projected clustering (as discussed below).  Since the contribution of shape noise to the galaxy-galaxy lensing covariance dominates over sample variance for scales $R < 10\,h^{-1}\,\mathrm{Mpc}$, the impact of non-linear galaxy bias and non-Gaussianity to the covariance is minimal \citep{2021A&A...646A.129J}.  The noise contribution to the covariance is determined by the intrinsic variation in the ellipticity measurements (shape noise, $\sigma_e$), and the effective source and lens number densities, $n_{\rm eff}$.  The survey overlap areas listed in Table~\ref{tab:SurveyCoverage} were used to determine the sample variance contribution to the galaxy-galaxy lensing covariance.  The full covariance matrix of the $\Delta\Sigma$ measurements has dimension $N_{z_s} N_{z_l} N_R$, where $N_{z_s}$ is the number of source samples, $N_{z_l}$ is the number of lens samples, and $N_R$ is the number of scale bins.

\subsection{Optimal compression of \texorpdfstring{$\Delta\Sigma$}{Delta Sigma} measurements}
\label{subsec:optCombDS}

Given that $\Delta\Sigma$ represents the projected mass density profile around the lenses, and does not vary with source sample (neglecting effects of intrinsic alignments and magnification which are not significant for our analysis), we compressed the set of $\Delta\Sigma$ measurements across different source tomographic samples for each lens bin, using the optimal data compression technique described in Appendix~C of \cite{blake_testing_2020}.  We apply this method by arranging the $\Delta\Sigma(R, z_s)$ measurements for each lens bin into a combined data vector $\mathbf{x}$ of length $N_{z_s}N_R$, with an associated covariance (sub)matrix $\mathbf{C}_x$ of dimension $N_{z_s}N_R \times N_{z_s}N_R$.  We then defined a weight matrix $\mathbf{w}$ of dimension $N_{z_s}N_R \times N_R$ to compress these measurements over the source dimension into a data vector $\mathbf{y} = \Delta\Sigma(R)$ of length $N_R$,
\begin{equation}
    \mathbf{y} = \mathbf{w}^{T}\,\mathbf{x} .
\end{equation}
The optimal weights that minimize the variance of the compressed measurement are given by,
\begin{equation}
    \mathbf{w} = \mathbf{C}_x^{-1}\,\mathbf{D} ,
\end{equation}
where \(\mathbf{D}\) is a boolean matrix that selects contributions corresponding to the same scale bin, with the normalization that the weights for each scale bin sum to unity.  The covariance matrix for the compressed data is then given by,
\begin{equation}
    \mathbf{C}_y = \mathbf{w}^{T}\,\mathbf{C}_x\,\mathbf{w} .
\end{equation}

\subsection{\texorpdfstring{$w_p(R)$}{w\_p(R)} measurements}
\label{subsec:wpMeasurements}

We measure the projected correlation function of the lens samples, $w_p(R)$, using the package \texttt{corrfunc}\footnote{\url{https://corrfunc.readthedocs.io/en/master/}} \citep{2020MNRAS.491.3022S}.  The DESI projected correlation function measurements are presented by \cite{2025arXiv250621677H}, and we use the same approach for our measurements from the BOSS and Buzzard catalogues.  We do not need to employ small-scale PIP (pairwise inverse probability) weights \citep{2025JCAP...04..074B} in our analysis, because we are only considering larger projected scales ($R > R_0$) for which their effect is negligible.  We note that the $w_p(R)$ measurements for the galaxies in the different redshift bins were made using the full footprint of the data in each case, not just the overlap regions with the lensing surveys, to maximize the signal-to-noise of these measurements.

The projected correlation function is estimated from $\xi_{gg}(R, \Pi)$, the two-dimensional galaxy-galaxy correlation function defined in terms of transverse separation $R$ and LoS separation $\Pi$. We estimate $\xi_{gg}$ from pair counts of data (D) and random (R) points \citep{1993ApJ...412...64L},
\begin{equation}
    \hat{\xi}(R,\Pi) = \frac{DD(R,\Pi)-2\,DR(R,\Pi)+RR(R,\Pi)}{RR(R,\Pi)} .
\end{equation}
We compute the pair counts using the same binning in $R$ as for the $\Delta\Sigma$ measurements, and $\Pi_\mathrm{max}=100\,h^{-1}\,\mathrm{Mpc}$.  The projected correlation function $w_p(R)$ then follows from integrating $\xi_{gg}(R,\Pi)$ along the LoS,
\begin{equation}
    w_p(R) = 2\,\sum_i \,\xi_{gg}(R,\,\Pi_i)\,\Delta\Pi_i,
\end{equation}
where we adopt $\Delta\Pi_i = 1\,h^{-1}\,\mathrm{Mpc}$ for $0<\Pi<100\,h^{-1}\,\mathrm{Mpc}$.

We determined the covariance matrix for the $w_p(R)$ measurements, and the cross-covariance between $\Delta\Sigma$ and $w_p$, using the Gaussian analytical covariance framework described by \cite{yuan_redshift_2024}.  The covariance matrices for $w_p$ have dimension $N_{z_l} N_R$.

\begin{figure}
    \centering
    \includegraphics[width=\columnwidth]{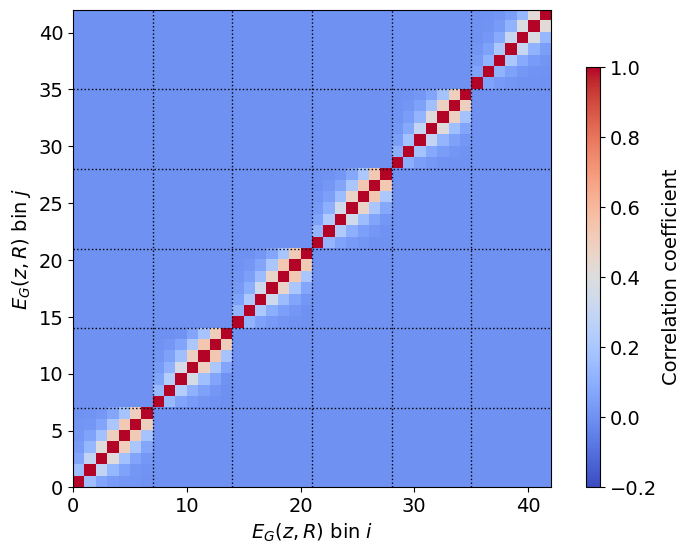}
    \caption{The correlation matrix $C_{ij}/\sqrt{C_{ii} C_{jj}}$ corresponding to the covariance matrix $C_{ij} = \text{Cov}[E^{l_i}_G(R_u),E^{l_j}_G(R_v)]$ given in Eq.~\ref{eq:egCovErr} for the joint analysis of the DESI-DR1 and KiDS-1000 datasets.  The $E_G(z,R)$ data vector is arranged looping first over redshift bins, then scale bins for $R > R_0$.  The dotted lines in the figure indicate the division into six lens redshift bins.}
    \label{fig:egcovariance}
\end{figure}

\subsection{\texorpdfstring{$\Upsilon$}{Upsilon} measurements}
\label{subsec:Y_Numerical}

Having defined our projected lensing ($\Delta \Sigma$) and clustering ($w_p$) observables, we now outline our determination of the annular statistics $\Upsilon_{gm}$ and $\Upsilon_{gg}$ introduced in Sec.~\ref{subsec:linearity}, including the propagation of the covariance into these observables.  These statistics suppress small-scale, highly non-linear contributions by subtracting the signal measured below a chosen cut-off radius $R_0$.  We discuss our selection of $R_0$ based on our analysis of the Buzzard mock catalogues in Sec.~\ref{sec:sim} below.

To compute $\Upsilon_{gm}$ from $\Delta \Sigma(R)$, we use Eq.~\ref{eq:Ygm_theory}. We first evaluate $\Delta \Sigma(R_0)$ by linear interpolation between the measured data points on either side of $R_0$, using error propagation to determine $\sigma[\Delta \Sigma(R_0)]$.  With $\Delta \Sigma(R_0)$ in hand, the annular galaxy-matter statistic is given by,
\begin{equation}
\label{eq:Y_gm_numerical}
    \Upsilon_{gm}(R) = \Delta \Sigma(R) - \frac{R_0^2}{R^2}\,\Delta \Sigma(R_0) .
\end{equation}
Propagating uncertainties from the $\Delta \Sigma$ covariance matrix yields the covariance of $\Upsilon_{gm}$,
\begin{equation}
\begin{split}
    & \text{Cov}[\Upsilon^{l_i}_{gm}(R_u), \Upsilon^{l_j}_{gm}(R_v)] = \\ & \text{Cov}[\Delta \Sigma^{l_i}(R_u), \Delta \Sigma^{l_j}(R_v)] + \frac{R^4_0}{R^2_u R^2_v}\sigma^2[\Delta \Sigma(R_0)],
\end{split}
\end{equation}
where $(l_i,l_j)$ denote lens redshift bins and $(R_u,R_v)$ denote radial bins.

The computation of $\Upsilon_{gg}(R)$ follows an analogous approach using Eq.~\ref{eq:Ygg_theory}. First, we determine $w_p(R_0)$ and $\sigma[w_p(R_0)]$ by interpolation.  The covariance of $\Upsilon_{gg}(R)$ can be obtained by re-writing Eq.~\ref{eq:Ygg_theory} in the form \citep[see][]{blake_testing_2020}, 
\begin{equation}
\label{eq:Ygg_numerical}
    \Upsilon_{gg}(R) = \frac{\rho_c}{R^2} \sum_{i=j}^{k} C_i\,w_{p,i} ,
\end{equation}
where $(j,k)$ are the indices of the $R$-bins containing $(R_0,R)$, and
\begin{equation}
    C_i = 
    \begin{cases}
        R_{i,\max}^2 & i=j, \\
        R_{i,\max}^2 - R_{i,\min}^2 & j<i<k, \\
        -\,R_{i,\min}^2 & i=k .
    \end{cases}
\end{equation}
We propagate errors through the mapping in Eq.~\ref{eq:Ygg_numerical} to derive the covariance matrix of $\Upsilon_{gg}(R)$ from that of $w_p(R)$,
\begin{equation}
\begin{split}
    &\text{Cov}[\Upsilon^{l_i}_{gg}(R_u),\,\Upsilon^{l_j}_{gg}(R_v)] 
    = \frac{\rho_c^2}{R_u^2 R_v^2} \sum_{u} \sum_{v}\, C_u\,C_v \\
    &\quad\, \times \text{Cov}[w_p^{l_i}(R_u),\,w_p^{l_j}(R_v)] .
\end{split}
\end{equation}

\subsection{\texorpdfstring{$E_G$}{E\_G} direct determination}
\label{subsec:EGdirect}

The estimate of $E_G$ also involves a determination of the RSD parameter $\beta$ for each lens sample, which we describe in Sec.~\ref{sec:sim} and Sec.~\ref{sec:egResults} depending on the context of the Buzzard, BOSS and DESI samples.  With this in place, we define the direct estimator for $E_G$ in lens redshift bin $l_i$ as,
\begin{equation}
\label{eq:EG_direct}
    E_{G}^{l_i}(R) = \frac{1}{\beta(z_i)} \,\frac{\Upsilon_{gm}^{l_i}(R)}{\Upsilon_{gg}^{l_i}(R)} .
\end{equation}
The covariance of $E_{G}^{l_i}(R)$ follows from propagating the errors in $\Upsilon_{gm}^{l_i}$, $\Upsilon_{gg}^{l_i}$ and $\beta$,
\begin{equation}
\label{eq:egCovErr}
\begin{split}
     &\frac{\text{Cov}[E^{l_i}_G(R_u) , E^{l_j}_G(R_v)]}{E^{l_i}_G(R_u) E^{l_j}_G(R_v)} = \frac{\text{Cov}[\Upsilon^{l_i}_{gm}(R_u), \Upsilon^{l_j}_{gm}(R_v)]}{\Upsilon^{l_i}_{gm}(R_u)\Upsilon^{l_j}_{gm}(R_v)} \\ &+ \frac{\text{Cov}[\Upsilon^{l_i}_{gg}(R_u), \Upsilon^{l_j}_{gg}(R_v)]}{\Upsilon^{l_i}_{gg}(R_u) \Upsilon^{l_j}_{gg}(R_v)} + \frac{\sigma_\beta^2}{\beta^2} ,
\end{split}
\end{equation}
which we assume are uncorrelated, given that the galaxy-galaxy lensing samples cover a small fraction of the area spanned by the clustering dataset.  An example correlation matrix for $\text{Cov}[E^{l_i}_G(R_u),E^{l_j}_G(R_v)]$ for the joint analysis of the DESI-DR1 and KiDS-1000 datasets is given in Fig.~\ref{fig:egcovariance}.

Under GR we expect $E_G$ to be scale-independent, so we further compress $E_{G}^{l_i}(R)$ across the separation bins $R$ within each lens bin $l_i$. Specifically, we apply the optimal weighting scheme described in Sec.~\ref{subsec:optCombDS} to obtain a single best estimate of the average $E_{G}^{l_i}$ per lens bin.

\subsection{\texorpdfstring{$E_G$}{E\_G} maximum-likelihood determination}
\label{subsec:egmaxlik}

The direct determination of $E_G$ utilises the ratio of noisy quantities $\Upsilon_{gm}(R)/\Upsilon_{gg}(R)$ across separation bins $R$, which may result in a biased or non-Gaussian result \citep{2023ApJS..267...21S, 2024OJAp....7E..97E}.  Therefore, following Sec 8.3 of \cite{blake_testing_2020}, we also performed a Bayesian maximum-likelihood fit for the average value $E_G(z)$ in each redshift slice.  In this case we performed a joint fit to $\Upsilon_{gm}(R)$ and $\Upsilon_{gg}(R)$ for each lens bin, where we vary a parameter $A_E$ which controls the amplitude of $\Upsilon_{gm}(R)$ relative to $\Upsilon_{gg}(R)$:
\begin{equation}
\label{eq:egmaxlik}
\begin{split}
    \Upsilon_{gm}(R) &= A_E \, b_L \, \Upsilon_{gm}(R, b_L, b_{NL}) , \\
    \Upsilon_{gg}(R) &= \Upsilon_{gg}(R, b_L, b_{NL}) .
\end{split}
\end{equation}
In this approach we also vary linear and non-linear galaxy bias parameters $b_L$ and $b_{NL}$, respectively, keeping all other cosmological parameters fixed at their fiducial values.  Hence we vary three parameters, $(A_E, b_L, b_{NL})$, for each lens redshift slice.\footnote{We note that the $b_L$ factor in the expression for $\Upsilon_{gm}$ in Eq.~\ref{eq:egmaxlik} ensures that $A_E$ is only constrained by the ratio $\Upsilon_{gm}(R)/\Upsilon_{gg}(R)$, given that $\Upsilon_{gm} \propto b_L$ and $\Upsilon_{gg} \propto b_L^2$, ``uncorrelating'' $A_E$ and $b_L$ in the fitting process and creating an estimator for $E_G$ which is a function of $A_E$ and $\beta$.}  We can then determine $E_G(z) = \Omega_m A_E/\beta$, after marginalising over the galaxy bias parameters \citep{blake_testing_2020}.

We evaluate the models $\Upsilon_{gm}(R, b_L, b_{NL})$ and $\Upsilon_{gg}(R, b_L, b_{NL})$ using non-linear galaxy-galaxy and galaxy-matter correlation functions \citep{2010PhRvD..81f3531B, 2013MNRAS.432.1544M}, assuming a local, non-linear galaxy bias relation in which the galaxy overdensity field is expressed in terms of the matter overdensity field as $\delta_g = b_L \delta_m + \frac{1}{2} b_{NL} \delta_m^2$.  These assumptions result in the correlation statistics \citep{2006PhRvD..74j3512M, 2009PhRvD..80f3528S},
\begin{equation}
\begin{split}
    \xi_{gg} &= b_L^2 \, \xi_{mm} + 2 \, b_L \, b_{NL} \, \xi_A + \frac{1}{2} b_{NL}^2 \, \xi_B, \\
    \xi_{gm} &= b_L \, \xi_{mm} + b_{NL} \, \xi_A ,
\end{split}
\end{equation}
where $\xi_{mm}$ is the matter correlation function, and the expressions for $\xi_A$ and $\xi_B$ are given in the references cited above.  We evaluated these functions using the \textsc{fast} software package \citep{2016JCAP...09..015M}.  We expect this model to only be valid on scales exceeding the virial radius of dark matter halos.

\subsection{Scale-dependent model}
\label{subsec:scaledep}

Together with the overall variation of the $E_G$ amplitude with redshift, we also tested the GR+$\Lambda$CDM prediction that the $E_G$ statistic should be independent of projected scale $R$, by fitting an empirical model across all redshift bins of the form \citep[following][]{blake_testing_2020},
\begin{equation}
    E_G(R, z_i) = A_i[1+\alpha \log_{10}(R)]\, ,
\label{eq:scaleModel}
\end{equation}
where $A_i$ is a free amplitude parameter in each redshift slice and $\alpha$ captures the logarithmic scale dependence of $E_G$. A value of $\alpha = 0$ corresponds to scale-independence, in which case $A_i = E_G(z_i)$.  Whilst Eq.~\ref{eq:scaleModel} is purely a phenomenological model, we note that predictions of $E_G$ in modified gravity scenarios have been presented by \cite{2015MNRAS.449.4326P} and \cite{2019JCAP...09..018P}.

\section{Simulation tests}
\label{sec:sim}

\begin{figure*}
    \centering
    \includegraphics[width=2\columnwidth]{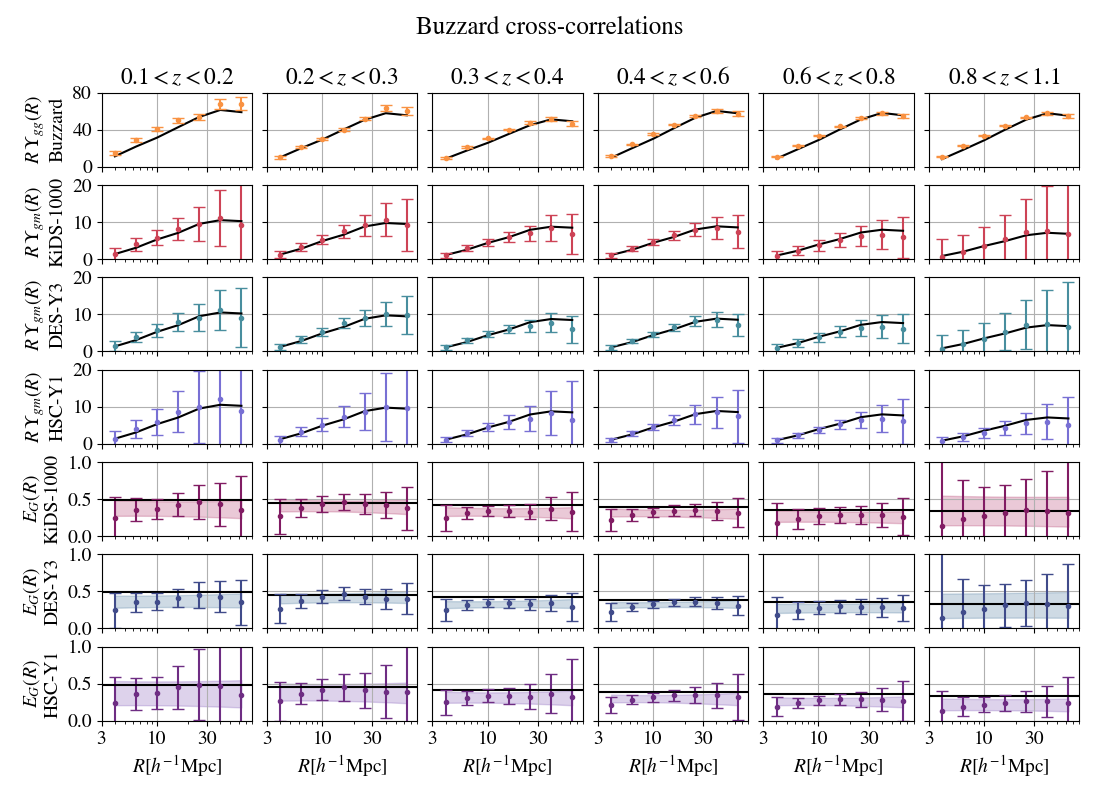}
    \caption{This figure displays mean cross-correlation measurements from the Buzzard simulation mock catalogues, assuming $R_0 = 3\, h^{-1}$ Mpc.  Each column represents one of the six simulated lens samples.  The top row shows $\Upsilon_{gg}(R)$ measurements, the next three rows show $\Upsilon_{gm}(R)$ measurements using these lens samples and the simulated KiDS-1000, DES-Y3 and HSC-Y1 source samples, and the final three rows show the corresponding $E_G(R,z)$ measurements using the direct ratio method.  The measurements are compared with the fiducial model prediction depicted by the solid black line, with errors from the diagonal of the covariance matrix.  The shaded regions in the $E_G(R,z)$ panels illustrate the 68\% confidence region of the fit of the general scale-dependent test model shown in Eq.~\ref{eq:scaleModel}.}
    \label{fig:eg_allBuzzard}
\end{figure*}

\begin{figure*}
    \centering
    \includegraphics[width=2\columnwidth]{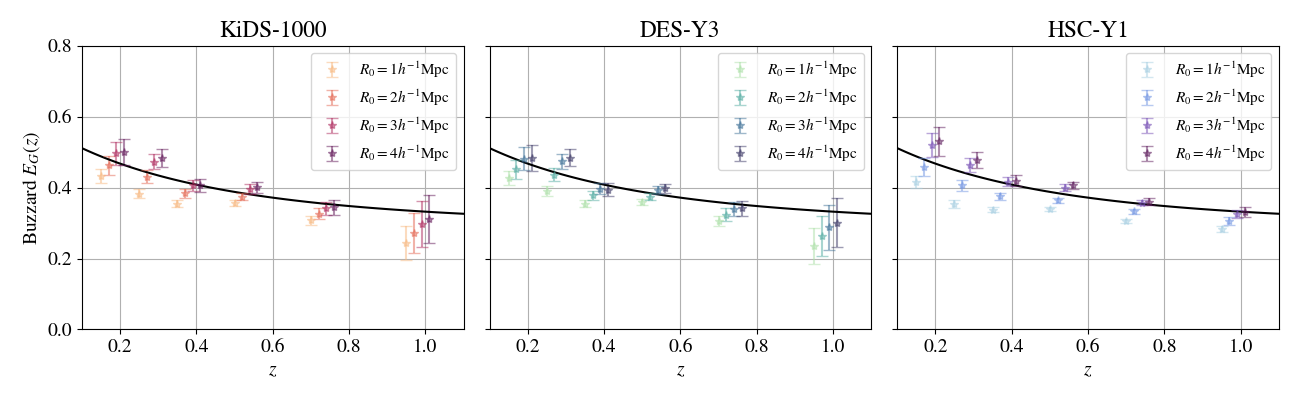}
    \caption{The determination of the scale-averaged value $E_G(z)$, using the maximum-likelihood method, for each lens redshift bin of the KiDS-1000, DES-Y3 and HSC-Y1 Buzzard simulations for varying non-linear cut-off scales $R_0 = (1, 2, 3, 4) \, h^{-1}$ Mpc.  The solid black line represents the fiducial model prediction.  The mean of the measurements across the $N_{\rm reg}$ mock regions is plotted, with the error scaled by $1/\sqrt{N_{\rm reg}}$.  The fiducial $E_G(z)$ values are recovered with the choice $R_0 = 3\, h^{-1}$ Mpc.}
\label{fig:egzsys}
\end{figure*}

\begin{figure*}
    \centering
    \includegraphics[width=2\columnwidth]{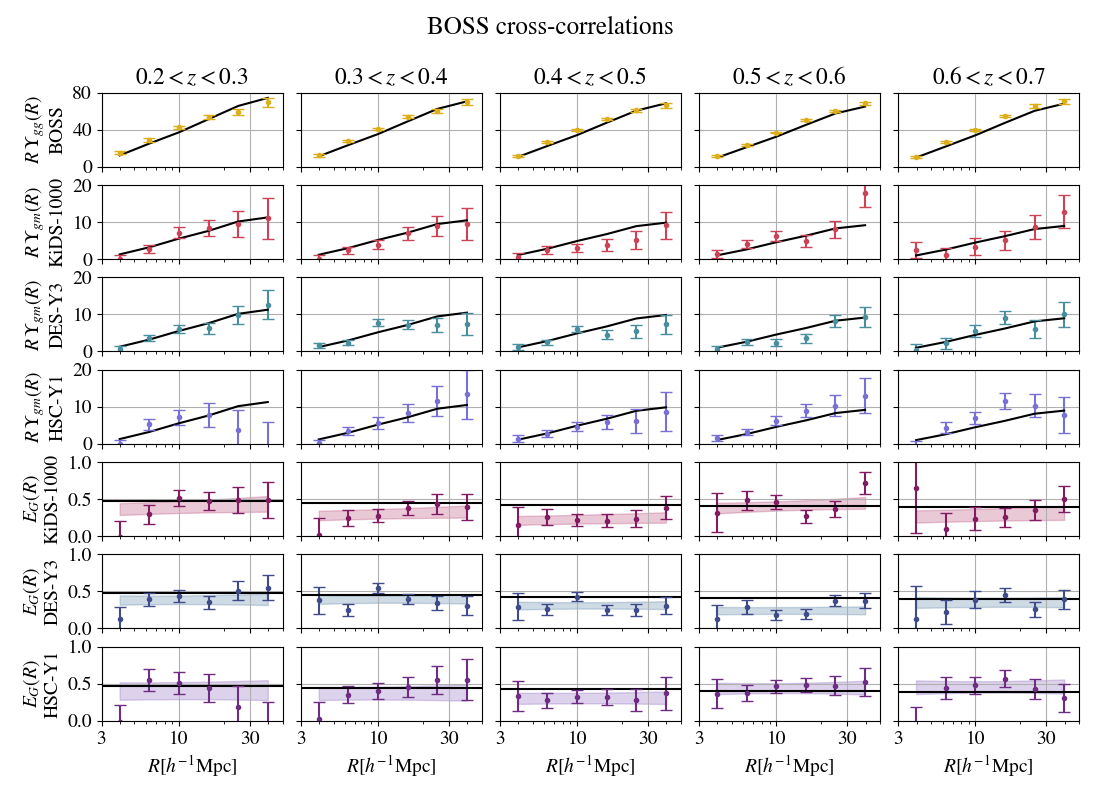}
    \caption{This figure displays cross-correlation measurements from the BOSS lens catalogues and KiDS-1000, DES-Y3 and HSC-Y1 source samples, displayed in the same style as Fig.~\ref{fig:eg_allBuzzard}.}
    \label{fig:eg_allBOSS}
\end{figure*}

In this section we describe the validation of our analysis pipeline using the Buzzard mock catalogues.  The measurements of $\Upsilon_{gg}(R)$ for the 6 simulated DESI lens samples, $\Upsilon_{gm}(R)$ for the three simulated weak lensing surveys KiDS-1000, DES-Y3 and HSC-Y1, and the corresponding $E_G(R,z)$ measurements by direct determination, are displayed in Fig.~\ref{fig:eg_allBuzzard}.   As discussed in Sec.~\ref{subsec:buzzard}, we cannot use the Buzzard simulations to perform a redshift-space distortion analysis.  Therefore, for the purposes of determining $E_G$ from the mocks, we assumed a fiducial value of $\beta = f_{\rm fid}/b_L$ by combining the fiducial growth rate $f_{\rm fid}$ with the best-fitting linear galaxy bias $b_L$ to the $\Upsilon_{gg}(R)$ measurements.  We assume $R_0 = 3\, h^{-1}$ Mpc for the measurements shown in Fig.~\ref{fig:eg_allBuzzard}, and we will validate this choice below.

The measurements shown in Fig.~\ref{fig:eg_allBuzzard} are averaged over the different mock regions accommodated in the Buzzard light cone (respectively $N_{\rm reg} =$ 20, 12 and 60 regions for the overlaps with KiDS, DES and HSC), the errors are derived from the analytical covariance representative of a single region, and the results are compared to the theoretical model predictions as defined in Sec.~\ref{sec:EGtheory}, using $\Omega_m = 0.286$ for the Buzzard simulations.  The mock measurements show generally good agreement with the model, except that the direct determinations of $E_G$ are consistently biased low compared to the model prediction.  \cite{alam_testing_2017} and \cite{singh_probing_2019} also find that $E_G$ measurements in simulations are biased low due to non-linear galaxy bias and a non-unity galaxy-matter cross-correlation coefficient, although their reported size of this effect is much smaller than the bias we find in our direct $E_G$ determination, which we attribute to the form of the estimator in Eq.~\ref{eq:EG_direct} as a ratio of noisy quantities.

As expected for these GR-consistent mocks, the $E_G$ measurements are broadly independent of scale.  In the $E_G(R,z)$ panels of Fig.~\ref{fig:eg_allBuzzard}, the shaded region indicates the 68\% confidence region of the fit of a general scale-dependent model introduced in Sec.~\ref{subsec:scaledep}.

The optimally-averaged $E_G(z)$ values, obtained by combining $E_G(R,z)$ across all separations using the full covariance matrix and the method described in Sec.~\ref{subsec:optCombDS}, are shown in the lower row of Fig.~\ref{fig:egzall}, in which we also divide the errors by $1/\sqrt{N_{\rm reg}}$.  These results confirm the impression that the direct estimator of $E_G$ is biased.  We also measured $E_G(z)$ for each mock lens redshift sample using the maximum-likelihood method described in Sec.~\ref{subsec:egmaxlik}, and display these results in Fig.~\ref{fig:egzsys} for different choices of non-linear cut-off scale $R_0 = (1, 2, 3, 4) \, h^{-1}$ Mpc.  Based on these results, we select $R_0 = 3 \, h^{-1}$ Mpc as a suitable choice providing an unbiased recovery of the theoretical value $E_G(z) = \Omega_m/f(z)$, and the maximum-likelihood determinations of $E_G(z)$ for this case are also displayed in the lower row of Fig.~\ref{fig:egzall} and listed in Table~\ref{tab:egresults}.  These results confirm that, using the maximum-likelihood method, the expected redshift dependence predicted by the underlying GR cosmology is reproduced, validating the robustness of the analysis pipeline.  For $R_0 = 3 \, h^{-1}$ Mpc, we find that the average deviation between the $E_G(z)$ measurements and model is consistent with zero with a precision of $2\%$, which is much smaller than the statistical errors in the measurements.  We note that our simulations are generated in the GR+$\Lambda$CDM framework, such that this non-linear threshold could potentially change in modified gravity scenarios.

\begin{table}
	\centering
	\caption{Measurements of $E_G(z)$ for different combinations of lens and source samples, performed using the maximum likelihood method and assuming $R_0 = 3 \, h^{-1}$ Mpc.}
	\label{tab:egresults}
	\begin{tabular}{c|c|c|c} 
		\hline 
		\textbf{Buzzard} & KiDS-1000 & DES-Y3 & HSC-Y1 \\ 
		\hline
            $E_G(\Bar{z} = 0.15)$ & $0.497 \pm 0.033$ & $0.482 \pm 0.033$ & $0.519 \pm 0.034$ \\
            $E_G(\Bar{z} = 0.25)$ & $0.473 \pm 0.022$ & $0.474 \pm 0.021$ & $0.464 \pm 0.020$ \\ 
            $E_G(\Bar{z} = 0.35)$ & $0.407 \pm 0.016$ & $0.397 \pm 0.016$ & $0.416 \pm 0.013$ \\
            $E_G(\Bar{z} = 0.5)$ & $0.396 \pm 0.013$ & $0.393 \pm 0.012$ & $0.400 \pm 0.009$ \\
            $E_G(\Bar{z} = 0.7)$ & $0.342 \pm 0.019$ & $0.339 \pm 0.019$ & $0.358 \pm 0.008$ \\
            $E_G(\Bar{z} = 0.95)$ & $0.297 \pm 0.064$ & $0.288 \pm 0.063$ & $0.325 \pm 0.012$ \\
		\hline
		\textbf{BOSS-DR12} & KiDS-1000 & DES-Y3 & HSC-Y1 \\ 
        \hline
            $E_G(\Bar{z} = 0.25)$ & $0.505 \pm 0.126$ & $0.485 \pm 0.103$ & $0.526 \pm 0.165$\\
            $E_G(\Bar{z} = 0.35)$ & $0.360 \pm 0.088$ & $0.464  \pm 0.079$ & $0.470 \pm 0.119$\\
            $E_G(\Bar{z} = 0.45)$ & $0.265 \pm 0.075$ & $0.354  \pm 0.064$ & $0.366 \pm 0.093$\\
            $E_G(\Bar{z} = 0.55)$ & $0.459 \pm 0.087$ & $0.276 \pm 0.060$ & $0.511 \pm 0.094$\\
            $E_G(\Bar{z} = 0.65)$ & $0.327 \pm 0.100$ & $0.386 \pm 0.084$ & $0.515 \pm 0.108$\\
        \hline
		\textbf{DESI-DR1} & KiDS-1000 & DES-Y3 & HSC-Y3 \\ 
        \hline
        $E_G(\Bar{z} = 0.15)$ & $0.368 \pm 0.192$ & $0.438 \pm 0.196$ & $0.568 \pm 0.225$  \\
        $E_G(\Bar{z} = 0.25)$ & $0.525 \pm 0.174$ & $0.587 \pm 0.184$ & $0.431 \pm 0.153$  \\
        $E_G(\Bar{z} = 0.35)$ & $0.447 \pm 0.143$ & $0.516 \pm 0.157$ & $0.561 \pm 0.167$  \\
        $E_G(\Bar{z} = 0.5)$ & $0.335 \pm 0.068$ & $0.253 \pm 0.055$ & $0.381 \pm 0.070$  \\
        $E_G(\Bar{z} = 0.7)$ & $0.287 \pm 0.064$ & $0.393 \pm 0.073$ & $0.345 \pm 0.055$  \\
        $E_G(\Bar{z} = 0.95)$ & $0.348 \pm 0.149$ & $0.218 \pm 0.123$ & $0.352 \pm 0.068$  \\
        \hline
    \end{tabular}
\end{table}

\section{Results}
\label{sec:egResults}

\subsection{Correlation measurements for BOSS-DR12}
\label{subsec:bosscorr}

We now present our correlation measurements using the BOSS-DR12 lens samples with the three weak lensing surveys KiDS-1000, DES-Y3 and HSC-Y1.  Fig.~\ref{fig:eg_allBOSS} presents the $\Upsilon_{gg}(R)$, $\Upsilon_{gm}(R)$ and corresponding $E_G(R,z)$ measurements by direct determination.  The errors are derived from the analytical covariance, and the results are compared to the theoretical model predictions using the best-fitting linear galaxy bias to the $\Upsilon_{gg}(R)$ measurements.

For the determination of $E_G$, we used the measurements of the RSD parameter $\beta$ for BOSS galaxies in narrow redshift slices by \cite{2019MNRAS.484..442Z}, following the method described in Sec 9.2 of \cite{blake_testing_2020}.  We again adopt $R_0 = 3 \, h^{-1}$ Mpc.  Fig.~\ref{fig:eg_allBOSS} compares the $E_G(R,z)$ measurements by direct determination to the predictions of the fiducial model $E_G(z) = \Omega_m/f(z)$ for the \textit{Planck} flat $\Lambda$CDM determination $\Omega_m = 0.3153$ \citep{aghanim_planck_2020}, and the best-fitting empirical scale-dependent model introduced in Eq.~\ref{eq:scaleModel}.  The middle row of Fig.~\ref{fig:egzall} compares the measurements of the scale-averaged $E_G(z)$ values at each redshift, using both direct determination and the maximum-likelihood method, and the latter results are also listed in Table~\ref{tab:egresults}.  In general, the $E_G(z)$ measurements using the maximum-likelihood method agree well with the \textit{Planck} model (which we discuss further in Sec.~\ref{subsec:reddep} below).  The best-fitting BOSS galaxy linear and non-linear bias parameters in the five lens redshift bins are: $b_L = (1.84 \pm 0.04, 1.93 \pm 0.03, 1.99 \pm 0.02, 2.10 \pm 0.02, 2.25 \pm 0.02)$ and $b_{NL} = (0.31 \pm 1.19, 0.28 \pm 0.16, 0.41 \pm 0.14, -0.06 \pm 0.16, 0.38 \pm 0.22)$.

Whilst our results are statistically consistent with previous $E_G(z)$ measurements using BOSS+2dFLenS+KiDS-1000 presented by \cite{blake_testing_2020}, there are minor methodological differences in the GGL measurements, modelling framework and source-lens cuts adopted by the two analyses.  Our current analysis presents the first measurements of $E_G$ from the combinations of BOSS+DES-Y3 and BOSS+HSC-Y1.

\subsection{Correlation measurements for DESI-DR1}
\label{subsec:desicorr}

\begin{figure*}
    \centering
    \includegraphics[width=2\columnwidth]{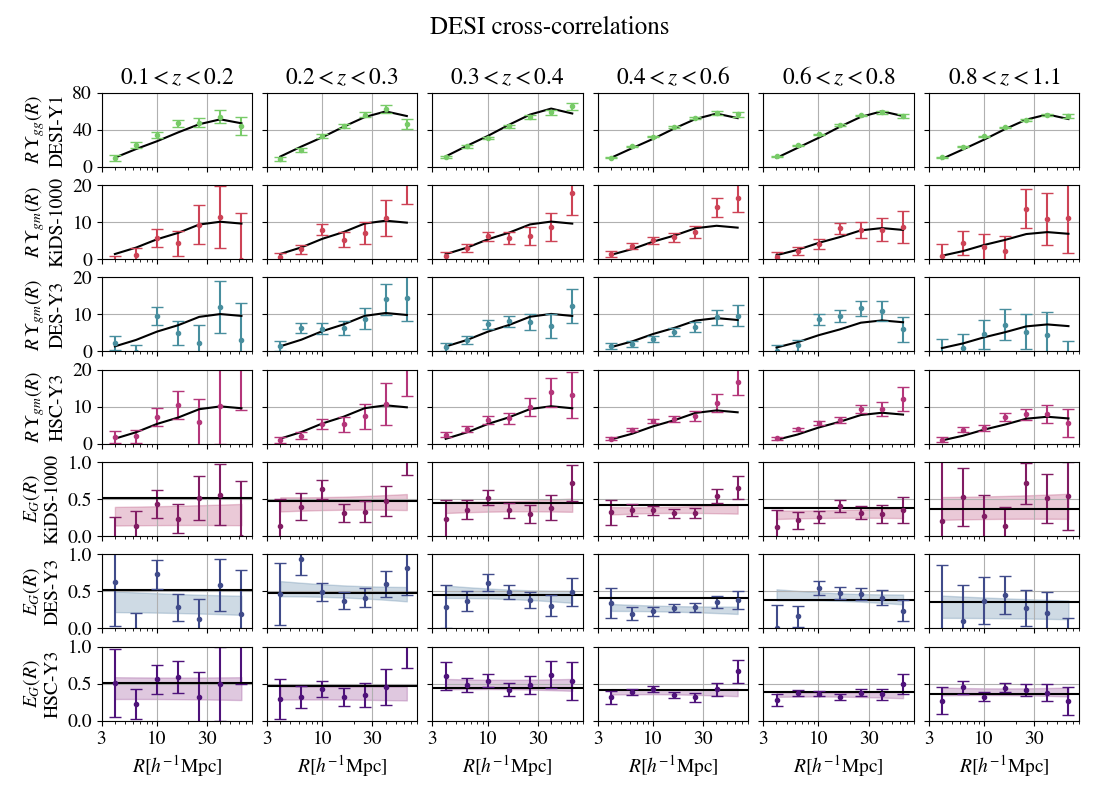}
    \caption{This figure displays cross-correlation measurements from the DESI-DR1 lens catalogues and KiDS-1000, DES-Y3 and HSC-Y3 source samples, displayed in the same style as Fig.~\ref{fig:eg_allBuzzard}.}
    \label{fig:eg_allDESI}
\end{figure*}

In this section, we present the correlation measurements using the six DESI-DR1 lens samples with the three weak lensing surveys KiDS-1000, DES-Y3 and HSC-Y3.\footnote{We note that the HSC-Y3 lensing catalogue became publicly available whilst finalising our DESI analysis, and we chose to use this dataset in preference to HSC-Y1 for the DESI results.  However, for completeness we also report results for DESI+HSC-Y1 in Table \ref{tab:all_eg_measurements}.}  Fig.~\ref{fig:eg_allDESI} presents the measurements of $\Upsilon_{gg}(R)$, $\Upsilon_{gm}(R)$ and $E_G(R,z)$ by direct determination.  The errors are derived from the analytical covariance, and the results are compared to the theoretical model predictions using the best-fitting linear galaxy bias to the $\Upsilon_{gg}(R)$ measurements.  We assume $R_0 = 3 \, h^{-1}$ Mpc throughout.

We used measurements of the RSD parameter $\beta$ for the DESI samples drawn from the Year 1 ``full-shape'' clustering analysis \citep{2025JCAP...09..008A}, where the parameter chains have been converted to $\beta$ values and errors by evaluating $f/b_L$ for each element of the chain.  The DESI collaboration have performed RSD measurements for LRGs in the same redshift bins as our analysis, and we find $\beta = (0.445 \pm 0.074, 0.448 \pm 0.064, 0.378 \pm 0.059)$ for redshift ranges $[0.4,0.6], [0.6,0.8], [0.8,1.1]$, respectively.  However, the full-shape analysis has not been performed in the narrower BGS redshift bins we use, only for the full $0.1 < z < 0.4$ dataset, producing $\beta = 0.380 \pm 0.103$ for this sample.  Hence, we apply this measurement to each of the BGS samples in our analysis, noting that the evolution with redshift is sub-dominant to the significant statistical error in the measurement of $\beta$.

Fig.~\ref{fig:eg_allDESI} displays the resulting $E_G(R,z)$ measurements by direct determination for each of the weak lensing datasets, compared with the fiducial $E_G$ model assuming the \textit{Planck} value $\Omega_m = 0.3153$.  As in the case of the BOSS analysis, we performed a scale-dependence test by fitting Eq.~\ref{eq:scaleModel} to the data vector, which includes a free overall amplitude at each redshift and a parameter $\alpha$ which determines the scale dependence, and these results are plotted as a 68\% confidence region.  The upper row of Fig.~\ref{fig:egzall} compares the measurements of the scale-averaged $E_G(z)$ values at each redshift, using both direct determination and the maximum-likelihood method, and the latter results are also listed in Table~\ref{tab:egresults}.  In the following sections we will interpret these results in terms of the redshift-dependent and scale-dependent predictions of the GR+$\Lambda$CDM model.  The best-fitting DESI galaxy linear and non-linear bias parameters in the six lens redshift bins are: $b_L = (1.24 \pm 0.10, 1.58 \pm 0.07, 1.77 \pm 0.02, 1.87 \pm 0.02, 2.09 \pm 0.02, 2.26 \pm 0.02)$ and $b_{NL} = (-2.14 \pm 1.16, -1.26 \pm 0.58, -0.01 \pm 0.24, 0.28 \pm 0.16, 0.51 \pm 0.16, 0.84 \pm 0.17)$.

\subsection{Redshift-dependence of \texorpdfstring{$E_G$}{E\_G}}
\label{subsec:reddep}

\begin{figure*}
    \centering
    \includegraphics[width=2\columnwidth]{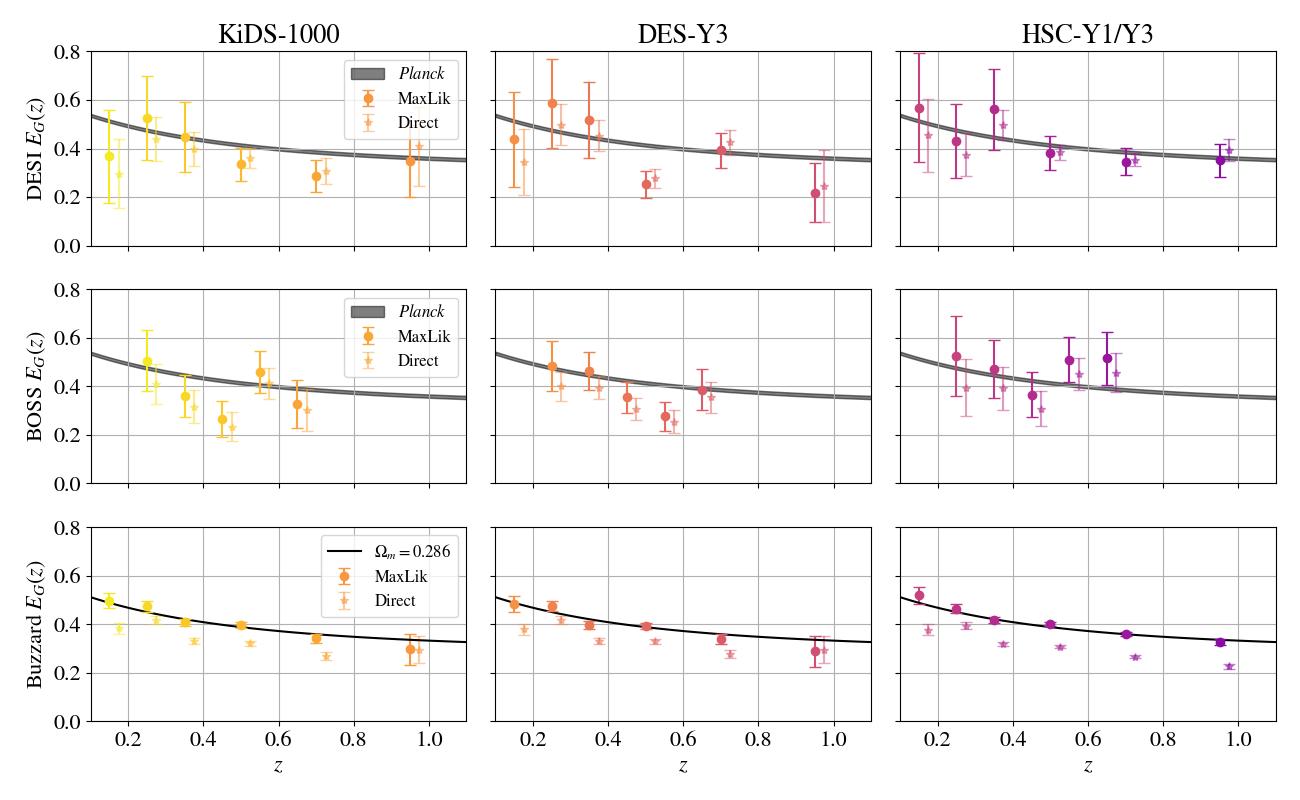}
    \caption{The scale-averaged $E_G(z)$ measurements for DESI (top row), BOSS (middle row), and Buzzard (bottom row), each combined with KiDS-1000 (left column), DES-Y3 (middle column) and HSC (right column), where HSC-Y3 is used for the DESI measurements, and HSC-Y1 is used for the BOSS and Buzzard measurements. The solid black line shows the $E_G(z)$ model $\Lambda$CDM prediction for the \textit{Planck} cosmology \citep{aghanim_planck_2020} in the upper two rows, where the width of the band indicates the $68\%$ confidence region, and the Buzzard fiducial cosmology in the bottom row.  Results are shown for the direct estimation and maximum likelihood methods for determining $E_G(z)$.}
    \label{fig:egzall} 
\end{figure*}

The GR+$\Lambda$CDM model provides a clear prediction for the dependence of the $E_G$ statistic on redshift, $E_G(z) = \Omega_m/f(z)$.  In this section we focus on the comparison of our $E_G$ measurements for BOSS-DR12 and DESI-DR1, in combination with KiDS-1000, DES-Y3, and HSC-Y1/Y3, with this predicted trend with redshift.

In Fig.~\ref{fig:egzall} we show the resulting determinations using the maximum-likelihood method of the average $E_G(z)$ as a function of redshift for DESI and BOSS, assuming a non-linear cut-off scale $R_0 = 3\,h^{-1}$ Mpc.  The solid black bands in the top and middle rows of Fig.~\ref{fig:egzall} indicate the GR+$\Lambda$CDM model prediction from \textit{Planck} satellite observations \citep{aghanim_planck_2020} with central value $\Omega_m = 0.3153$, where the width of the band indicates the $68\%$ confidence region.  The $\chi^2$ values of the predicted model with $\Omega_m = 0.3153$ against the maximum likelihood $E_G$ determinations are $(4.4, 10.8, 1.4)$ for DESI-DR1 using (KiDS-1000, DES-Y3, HSC-Y3), respectively, for 6 degrees of freedom (i.e., no fitted parameters).  These $\chi^2$ values all lie within likely ranges: the $\chi^2$ distribution for 6 degrees of freedom has a mean and standard deviation $6 \pm 3.5$.  For BOSS-DR12 measurements, the $\chi^2$ values of this model are $(6.4, 6.0, 3.1)$ using (KiDS-1000, DES-Y3, HSC-Y1), for 5 degrees of freedom, also indicating that the models provide a good fit to the data.  In all cases, the measurements are consistent with the \textit{Planck} prediction of the standard GR+$\Lambda$CDM model.

\subsection{Scale-dependence of \texorpdfstring{$E_G$}{E\_G}}

\begin{figure}
    \centering
    \includegraphics[width=\columnwidth]{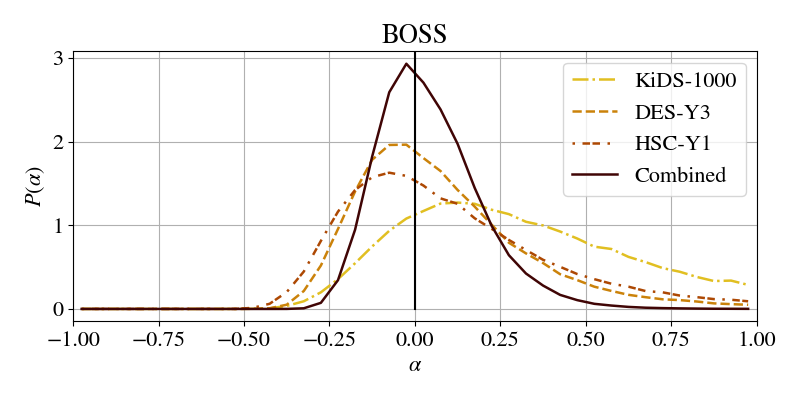}
    \includegraphics[width=\columnwidth]{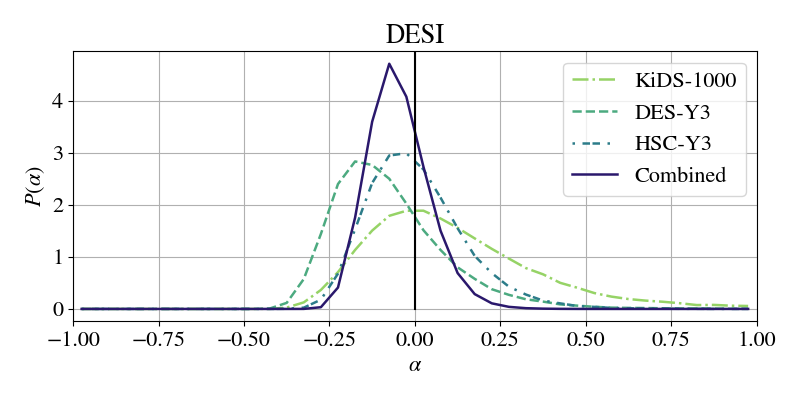}
    \caption{The posterior probability distributions $P(\alpha)$ resulting from the test for scale-dependence in $E_G(R,z)$ for the BOSS and DESI samples described by the test model $E_G(R,z_i) = A_i[1 + \alpha \log_{10}(R)]$, where $\alpha = 0$ defines the scale-independent case.  The broken lines represent $P(\alpha)$ for each distinct source sample, and the solid line shows the combination of these posterior distributions for the different source surveys.}
    \label{fig:scaleTests}
\end{figure}

Finally, we report our results for the fits of the phenomenological scale-dependent model of $E_G$ introduced in Sec.~\ref{subsec:scaledep}, in terms of the free parameter $\alpha$, where the GR+$\Lambda$CDM prediction is that the $E_G$ statistic should be independent of scale $R$ (i.e., $\alpha = 0$).  Fig.~\ref{fig:scaleTests} shows the posterior probability distributions of $\alpha$ for each weak lensing survey when combined with BOSS (top) and DESI (bottom). In all cases, $\alpha = 0$ lies well within the credible intervals, indicating no significant detection of scale dependence.  If we combine the measurements for different weak lensing surveys assuming they are statistically independent \citep[e.g.,][]{2023OJAp....6E..36D}, we find 68\% confidence regions $\alpha = 0.04 \pm 0.14$ for BOSS, and $\alpha = -0.05 \pm 0.09$ for DESI.  The significant improvement from the $\alpha = 0.17 \pm 0.26$ reported by \cite{blake_testing_2020} for BOSS+2dFLenS+KiDS-1000 is due to the additional weak lensing surveys included and improved constraining power of DESI LRGs.

\section{Discussion}
\label{sec:discussion}

The DESI observing strategy has allowed us to survey a wider range of redshifts for galaxy-galaxy lensing than previous studies, allowing us to present the lowest-redshift and highest-redshift measurements of $E_G(z)$ acquired at the time of writing (at $z=0.15$ and $z=0.95$).  A reduction in $E_G$ errors is evident for the DESI LRG measurements compared to similar redshift bins in BOSS, such that the DESI measurements are the most precise $E_G(z)$ measurements to date.  This enhanced precision arises from an increase in the number density of lenses.  Compared to BOSS, DESI has been able to collect a total of up to 8 times more source-lens pairs in the KiDS-1000 survey, 3.5 times more in DES-Y3 and 4.5 times as many in HSC-Y1.

The DESI BGS $E_G(z)$ measurements have larger errors than resulting from the comparable BOSS sample, owing to the reduced lens number density implied by the absolute magnitude cut $M_r < -21.5$ applied by the DESI Key Project and the consequent larger uncertainty in $\beta$.  The $E_G(z)$ measurements in the BGS redshift range could be improved by extending the analysis to higher number densities than currently performed by the collaboration.  Nevertheless, our work has demonstrated consistency in the determination of $E_G(z)$ between the BOSS and DESI datasets.

We tested the impact of systematic errors in the source redshift distributions of each tomographic bin, as calibrated by the weak lensing collaborations, on our $E_G(z)$ measurements.  We propagated the error in the offset of the mean redshift $\Delta z$ of each source distribution from its fiducial form \citep[using values from][]{2021A&A...647A.124H, 2022PhRvD.105b3520A, 2023PhRvD.108l3518L} into a shift of the resulting $E_G(z)$ amplitude, relative to its statistical error.  In all cases this offset is sub-dominant to the statistical error in $E_G(z)$ and, except for the highest two DESI lens redshift bins for HSC-Y3 sources, the shift is $<0.2 \sigma$.  For the $z = 0.7$ and $z = 0.85$ DESI redshift bins for HSC-Y3, we find an additional error in $E_G(z)$ of $0.3-0.4 \sigma$.  We conclude that whilst this source of systematic error is currently not significant for our work, as these measurements improve in accuracy it will be important to consider the redshift distributions of the highest source bins in HSC-Y3.

\begin{figure*}
    \centering
    \includegraphics[width=2\columnwidth]{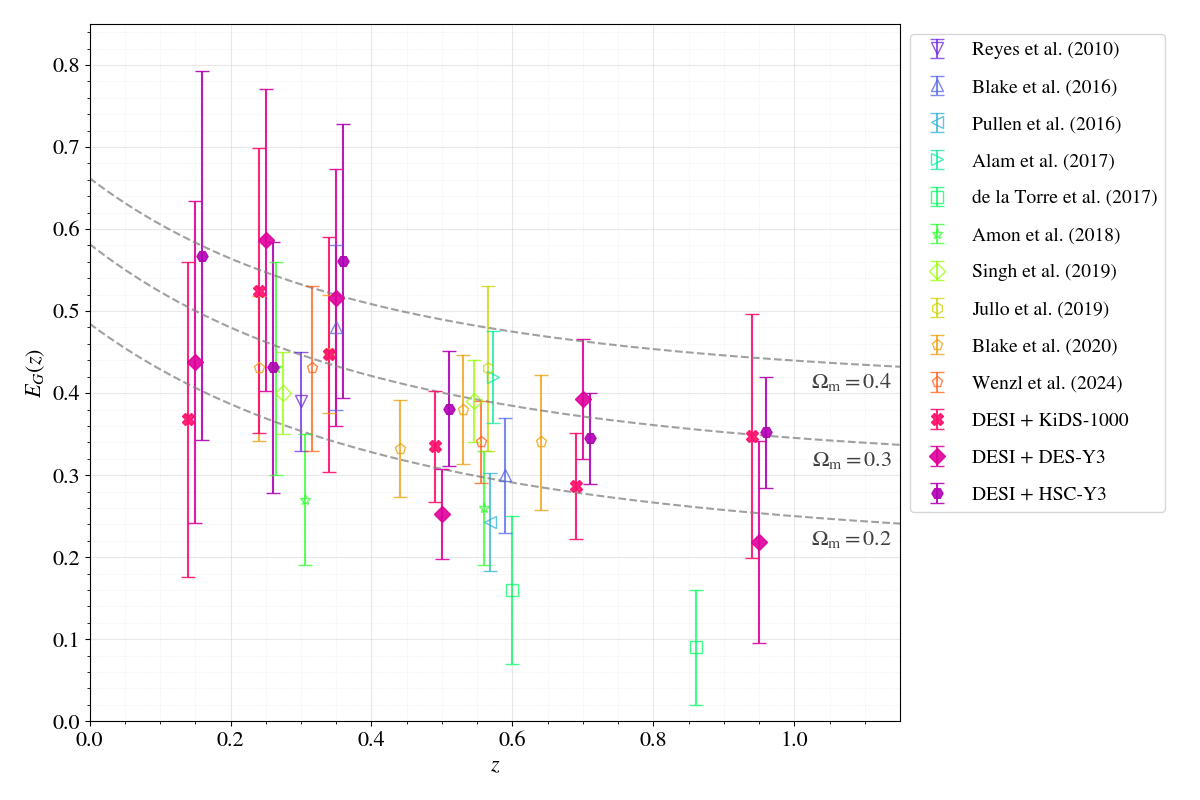}
    \caption{A summary of current $E_G(z)$ measurements, combining pre-existing literature and our new DESI measurements.  For clarity of presentation, we have only included the new results derived from DESI, not BOSS.  A full set of tabulated data can be found in Table~\ref{tab:all_eg_measurements}.}
    \label{fig:finalEG_all}
\end{figure*}

We have considered the redshift dependence and scale dependence of the $E_G$ measurements as a test of the GR+$\Lambda$CDM model.  We do not detect any measurable scale dependence in the results, validating the GR model prediction.  Regarding redshift dependence: whilst some previous measurements have reported an $E_G$ amplitude somewhat lower than the model prediction \citep[as discussed by][]{skara_tension_2020}, our results are statistically consistent with the predictions of the \textit{Planck} fiducial cosmology, which is the displayed as the solid black line in the BOSS and DESI panels of Fig.~\ref{fig:egzall}.  Fig.~\ref{fig:finalEG_all} provides a snapshot of all the $E_G$ literature measurements over a 15-year period to date, including our new contributions.  The dashed lines are reference models for three different $\Omega_m$ values.  A list of individual $E_G(z)$ measurements from galaxy weak lensing is also provided in Table~\ref{tab:all_eg_measurements} in Appendix~\ref{sec:eg_history}.

\section{Conclusions}
\label{sec:conclusion}

Our results have highlighted the role that large redshift surveys can play, in conjunction with powerful weak lensing datasets, in creating precise cosmological measurements and constraining gravity on large scales.  In this study we have focussed on performing new consistency tests of the GR+$\Lambda$CDM model calibrated by \textit{Planck} using the gravitational estimator statistic $E_G$, which compares the amplitudes of light deflection via gravitational lensing, with the amplitude of galaxy velocities traced using redshift-space distortions.  We have validated our analysis pipeline using cosmological simulations.

In closing, we highlight three key conclusions from our research:
\begin{itemize}
    \item Our $E_G(R,z)$ measurements match the scale-independent prediction of GR.
    \item The amplitude of $E_G(z)$ matches the predictions of the standard GR+$\Lambda$CDM model in the \textit{Planck} fiducial cosmology, which we have tested over an unprecedented range of redshifts.
    \item The increasing precision of the measurements of $E_G(z)$, now approaching $10\%$, necessitates careful consideration of systematic error contributions, such as those associated with non-linear galaxy bias, intrinsic alignments and lens magnification.
\end{itemize}
Our study using the $E_G$ statistic is complementary to current and upcoming analyses of these datasets which adopt phenomenological modified gravity frameworks \citep{2024arXiv241112026I, CristhianDESI}.

\section*{Acknowledgements}

We thank Ginevra Favole and Rachel Bean, and two anonymous journal reviewers, for providing useful comments on a draft of this paper.

SR would like to acknowledge the financial support received through the award of a Research Training Program Stipend scholarship by Swinburne University.  CB acknowledges financial support received through Australian Research Council Discovery Project DP220101609.  UA acknowledges support from the Leinweber Center for Theoretical Physics at the University of Michigan Postdoctoral Research Fellowship and DOE grant DE-FG02-95ER40899, and HN acknowledges support by SECIHTI grant CBF2023-2024-162, PAPIIT IA101825, and PAPIIT IN101124.

This material is based upon work supported by the U.S. Department of Energy (DOE), Office of Science, Office of High-Energy Physics, under Contract No. DE–AC02–05CH11231, and by the National Energy Research Scientific Computing Center, a DOE Office of Science User Facility under the same contract. Additional support for DESI was provided by the U.S. National Science Foundation (NSF), Division of Astronomical Sciences under Contract No. AST-0950945 to the NSF's National Optical-Infrared Astronomy Research Laboratory; the Science and Technology Facilities Council of the United Kingdom; the Gordon and Betty Moore Foundation; the Heising-Simons Foundation; the French Alternative Energies and Atomic Energy Commission (CEA); the National Council of Humanities, Science and Technology of Mexico (CONAHCYT); the Ministry of Science, Innovation and Universities of Spain (MICIU/AEI/10.13039/501100011033), and by the DESI Member Institutions: \url{https://www.desi.lbl.gov/collaborating-institutions}. Any opinions, findings, and conclusions or recommendations expressed in this material are those of the author(s) and do not necessarily reflect the views of the U.S. National Science Foundation, the U.S. Department of Energy, or any of the listed funding agencies.

The authors are honored to be permitted to conduct scientific research on I'oligam Du'ag (Kitt Peak), a mountain with particular significance to the Tohono O’odham Nation.

\section*{Data availability}

Data points for the figures are available at \url{https://doi.org/10.5281/zenodo.15726063}.

\bibliographystyle{mnras}
\bibliography{Lensing}

\section*{Affiliations}
\scriptsize
\noindent
$^{1}$ Centre for Astrophysics \& Supercomputing, Swinburne University of Technology, P.O. Box 218, Hawthorn, VIC 3122, Australia\\
$^{2}$ Leinweber Center for Theoretical Physics, University of Michigan, 450 Church Street, Ann Arbor, Michigan 48109-1040, USA\\
$^{3}$ University of Michigan, 500 S. State Street, Ann Arbor, MI 48109, USA\\
$^{4}$ Instituto de Ciencias F\'{\i}sicas, Universidad Nacional Aut\'onoma de M\'exico, Av. Universidad s/n, Cuernavaca, Morelos, C.~P.~62210, M\'exico\\
$^{5}$ Instituto de F\'{\i}sica, Universidad Nacional Aut\'{o}noma de M\'{e}xico,  Circuito de la Investigaci\'{o}n Cient\'{\i}fica, Ciudad Universitaria, Cd. de M\'{e}xico  C.~P.~04510,  M\'{e}xico\\
$^{6}$ Lawrence Berkeley National Laboratory, 1 Cyclotron Road, Berkeley, CA 94720, USA\\
$^{7}$ Department of Physics, Boston University, 590 Commonwealth Avenue, Boston, MA 02215 USA\\
$^{8}$ Department of Physics \& Astronomy, University of Rochester, 206 Bausch and Lomb Hall, P.O. Box 270171, Rochester, NY 14627-0171, USA\\
$^{9}$ Dipartimento di Fisica ``Aldo Pontremoli'', Universit\`a degli Studi di Milano, Via Celoria 16, I-20133 Milano, Italy\\
$^{10}$ INAF-Osservatorio Astronomico di Brera, Via Brera 28, 20122 Milano, Italy\\
$^{11}$ Department of Physics \& Astronomy, University College London, Gower Street, London, WC1E 6BT, UK\\
$^{12}$ Physics Department, Brookhaven National Laboratory, Upton, NY 11973, USA\\
$^{13}$ University of California, Berkeley, 110 Sproul Hall \#5800 Berkeley, CA 94720, USA\\
$^{14}$ Departamento de F\'isica, Universidad de los Andes, Cra. 1 No. 18A-10, Edificio Ip, CP 111711, Bogot\'a, Colombia\\
$^{15}$ Observatorio Astron\'omico, Universidad de los Andes, Cra. 1 No. 18A-10, Edificio H, CP 111711 Bogot\'a, Colombia\\
$^{16}$ Center for Astrophysics $|$ Harvard \& Smithsonian, 60 Garden Street, Cambridge, MA 02138, USA\\
$^{17}$ Institut d'Estudis Espacials de Catalunya (IEEC), c/ Esteve Terradas 1, Edifici RDIT, Campus PMT-UPC, 08860 Castelldefels, Spain\\
$^{18}$ Institute of Cosmology and Gravitation, University of Portsmouth, Dennis Sciama Building, Portsmouth, PO1 3FX, UK\\
$^{19}$ Institute of Space Sciences, ICE-CSIC, Campus UAB, Carrer de Can Magrans s/n, 08913 Bellaterra, Barcelona, Spain\\
$^{20}$ Fermi National Accelerator Laboratory, PO Box 500, Batavia, IL 60510, USA\\
$^{21}$ Department of Astronomy and Astrophysics, UCO/Lick Observatory, University of California, 1156 High Street, Santa Cruz, CA 95064, USA\\
$^{22}$ Center for Cosmology and AstroParticle Physics, The Ohio State University, 191 West Woodruff Avenue, Columbus, OH 43210, USA\\
$^{23}$ Department of Physics, The Ohio State University, 191 West Woodruff Avenue, Columbus, OH 43210, USA\\
$^{24}$ The Ohio State University, Columbus, 43210 OH, USA\\
$^{25}$ School of Mathematics and Physics, University of Queensland, Brisbane, QLD 4072, Australia\\
$^{26}$ Department of Physics, University of Michigan, 450 Church Street, Ann Arbor, MI 48109, USA\\
$^{27}$ Department of Physics, The University of Texas at Dallas, 800 W. Campbell Rd., Richardson, TX 75080, USA\\
$^{28}$ CIEMAT, Avenida Complutense 40, E-28040 Madrid, Spain\\
$^{29}$ NSF NOIRLab, 950 N. Cherry Ave., Tucson, AZ 85719, USA\\
$^{30}$ Aix Marseille Univ, CNRS, CNES, LAM, Marseille, France\\
$^{31}$ Department of Physics, Southern Methodist University, 3215 Daniel Avenue, Dallas, TX 75275, USA\\
$^{32}$ Department of Physics and Astronomy, University of California, Irvine, 92697, USA\\
$^{33}$ Department of Physics and Astronomy, University of Waterloo, 200 University Ave W, Waterloo, ON N2L 3G1, Canada\\
$^{34}$ Perimeter Institute for Theoretical Physics, 31 Caroline St. North, Waterloo, ON N2L 2Y5, Canada\\
$^{35}$ Waterloo Centre for Astrophysics, University of Waterloo, 200 University Ave W, Waterloo, ON N2L 3G1, Canada\\
$^{36}$ Department of Physics, American University, 4400 Massachusetts Avenue NW, Washington, DC 20016, USA\\
$^{37}$ Sorbonne Universit\'{e}, CNRS/IN2P3, Laboratoire de Physique Nucl\'{e}aire et de Hautes Energies (LPNHE), FR-75005 Paris, France\\
$^{38}$ Department of Astronomy and Astrophysics, University of California, Santa Cruz, 1156 High Street, Santa Cruz, CA 95065, USA\\
$^{39}$ Departament de F\'{i}sica, Serra H\'{u}nter, Universitat Aut\`{o}noma de Barcelona, 08193 Bellaterra (Barcelona), Spain\\
$^{40}$ Institut de F\'{i}sica d’Altes Energies (IFAE), The Barcelona Institute of Science and Technology, Edifici Cn, Campus UAB, 08193, Bellaterra (Barcelona), Spain\\
$^{41}$ Instituci\'{o} Catalana de Recerca i Estudis Avan\c{c}ats, Passeig de Llu\'{\i}s Companys, 23, 08010 Barcelona, Spain\\
$^{42}$ Department of Physics \& Astronomy and Pittsburgh Particle Physics, Astrophysics, and Cosmology Center (PITT PACC), University of Pittsburgh, 3941 O'Hara Street, Pittsburgh, PA 15260, USA\\
$^{43}$ Departamento de F\'{\i}sica, DCI-Campus Le\'{o}n, Universidad de Guanajuato, Loma del Bosque 103, Le\'{o}n, Guanajuato C.~P.~37150, M\'{e}xico\\
$^{44}$ Instituto Avanzado de Cosmolog\'{\i}a A.~C., San Marcos 11 - Atenas 202. Magdalena Contreras. Ciudad de M\'{e}xico C.~P.~10720, M\'{e}xico\\
$^{45}$ IRFU, CEA, Universit\'{e} Paris-Saclay, F-91191 Gif-sur-Yvette, France\\
$^{46}$ Institute for Astronomy, University of Edinburgh, Royal Observatory, Blackford Hill, Edinburgh EH9 3HJ, UK\\
$^{47}$ Ruhr University Bochum, Faculty of Physics and Astronomy, Astronomical Institute (AIRUB), German Centre for Cosmological Lensing, 44780 Bochum, Germany\\
$^{48}$ Instituto de Astrof\'{i}sica de Andaluc\'{i}a (CSIC), Glorieta de la Astronom\'{i}a, s/n, E-18008 Granada, Spain\\
$^{49}$ Departament de F\'isica, EEBE, Universitat Polit\`ecnica de Catalunya, c/Eduard Maristany 10, 08930 Barcelona, Spain\\
$^{50}$ Department of Physics and Astronomy, Sejong University, 209 Neungdong-ro, Gwangjin-gu, Seoul 05006, Republic of Korea\\
$^{51}$ Queensland University of Technology,  School of Chemistry \& Physics, George St, Brisbane 4001, Australia\\
$^{52}$ Max Planck Institute for Extraterrestrial Physics, Gie\ss enbachstra\ss e 1, 85748 Garching, Germany\\
$^{53}$ Department of Astronomy, Tsinghua University, 30 Shuangqing Road, Haidian District, Beijing, China, 100190\\
$^{54}$ National Astronomical Observatories, Chinese Academy of Sciences, A20 Datun Road, Chaoyang District, Beijing, 100101, P.~R.~China\\
\normalsize

\appendix
\section{History of galaxy weak lensing \texorpdfstring{$E_G$}{E\_G} measurements}
\label{sec:eg_history}

This section contains a list of galaxy weak lensing $E_G$ measurements published to date, sorted by increasing redshift.

\begingroup
\tiny
\begin{longtable}{|c|cccc|c|}
\caption{History of galaxy weak lensing $E_G$ measurements}
\label{tab:all_eg_measurements} \\
\cline{2-6}
     \multicolumn{1}{c|}{\textbf{No.}} & \centering\arraybackslash\textbf{Dataset}  & \centering\arraybackslash{$\boldsymbol{\Delta z}$} & \centering\arraybackslash{$\boldsymbol{E_G(\Bar{z})}$} & \centering\arraybackslash{$\boldsymbol{R\ [h^{-1}\, \textbf{Mpc}]}$} &\multicolumn{1}{c|}{\textbf{Publication}} \\
\hline
        & & & & & \\
\endfirsthead

\multicolumn{6}{c}%
{{\textbf{Table \thetable:} Continued from previous page.}} \\
\cline{2-6}
     \multicolumn{1}{c|}{\textbf{No.}} & \centering\arraybackslash\textbf{Dataset}  & \centering\arraybackslash{$\boldsymbol{\Delta z}$} & \centering\arraybackslash{$\boldsymbol{E_G(\Bar{z})}$} & \centering\arraybackslash{$\boldsymbol{R\ [h^{-1}\, \textbf{Mpc}]}$} &\multicolumn{1}{c|}{\textbf{Publication}} \\
\hline
        & & & & & \\
\endhead
        & & & & & \\
\hline \multicolumn{6}{r}{{Continued on next page}} 
\endfoot
\hline
\endlastfoot
   
      1 & KiDS-1000+DESI-DR1 & $0.1<z<0.2$ & $0.37 \pm 0.19$ & $3<R<80$ & \textit{Current}   \\
      2 & DES-Y3+DESI-DR1 & $0.1<z<0.2$ & $0.44 \pm 0.20$ & $3<R<80$ & \textit{Current}   \\
      3 & HSC-Y1+DESI-DR1 & $0.1<z<0.2$ & $0.45 \pm 0.27$ & $3<R<80$ & \textit{Current}   \\
      4 & HSC-Y3+DESI-DR1 & $0.1<z<0.2$ & $0.57 \pm 0.23$ & $3<R<80$ & \textit{Current}   \\
       & & & & & \\
      5 & RCSLenS+CFHTLenS+WGZLoZ+LOWZ &  $0.15<z<0.43$ & $0.40 \pm 0.09 $ & $3<R<50$ &  \cite{blake_rcslens_2016} \\
      6 & KiDS-450+LOWZ+2dFLOZ & $0.15<z<0.43$ & $0.27 \pm 0.08$ & $5<R<60$ &  \cite{amon_kids2dflensgama_2018} \\
      7 & KiDS-450+GAMA  & $0.15<z<0.51$ & $0.43 \pm 0.13$ & $5<R<40$ & \cite{amon_kids2dflensgama_2018} \\
      8 & SDSS+LOWZ  & $0.16<z<0.36$ & $0.40 \pm 0.042$ & $25<R<150$ & \cite{singh_probing_2019}  \\
      9 & SDSS  & $0.16<z<0.47$ & $0.39 \pm 0.065  $ & $10<R<50$ & \cite{reyes_confirmation_2010}  \\
       & & & & & \\
      10 & KiDS-1000+BOSS-DR12+2dFLenS & $0.2<z<0.3$ & $0.43 \pm 0.09$ & $3<R<100$ & \cite{blake_testing_2020}  \\
      11 & KiDS-1000+BOSS-DR12 & $0.2<z<0.3$ & $0.51 \pm 0.13$ & $3<R<50$ &  \textit{Current} \\
      12 & DES-Y3+BOSS-DR12 & $0.2<z<0.3$ & $0.49 \pm 0.10$ & $3<R<50$ & \textit{Current}   \\
      13 & HSC-Y1+BOSS-DR12 & $0.2<z<0.3$ & $0.53 \pm 0.17$ & $3<R<50$ &  \textit{Current}  \\
      14 & KiDS-1000+DESI-DR1 & $0.2<z<0.3$ & $0.53 \pm 0.17$ & $3<R<80$ & \textit{Current}   \\
      15 & DES-Y3+DESI-DR1 & $0.2<z<0.3$ & $0.59 \pm 0.18$ & $3<R<80$ & \textit{Current}   \\
      16 & HSC-Y1+DESI-DR1 & $0.2<z<0.3$ & $0.27 \pm 0.17$ & $3<R<80$ & \textit{Current}   \\
      17 & HSC-Y3+DESI-DR1 & $0.2<z<0.3$ & $0.43 \pm 0.15$ & $3<R<80$ & \textit{Current}   \\
         & & & & & \\
      18 & KiDS-1000+BOSS-DR12+2dFLenS & $0.3<z<0.4$ & $0.45 \pm 0.07$ & $3<R<100$ & \cite{blake_testing_2020}  \\
      19 & KiDS-1000+BOSS-DR12 & $0.3<z<0.4$ & $0.36 \pm 0.09$ & $3<R<50$ &  \textit{Current} \\
      20 & DES-Y3+BOSS-DR12 & $0.3<z<0.4$ & $0.46 \pm 0.08$ & $3<R<50$ & \textit{Current}   \\
      21 & HSC-Y1+BOSS-DR12 & $0.3<z<0.4$ & $0.47 \pm 0.12$ & $3<R<50$ &  \textit{Current}  \\
      22 & KiDS-1000+DESI-DR1 & $0.3<z<0.4$ & $0.45 \pm 0.14$ & $3<R<80$ & \textit{Current}   \\
      23 & DES-Y3+DESI-DR1 & $0.3<z<0.4$ & $0.52 \pm 0.16$ & $3<R<80$ & \textit{Current}   \\
      24 & HSC-Y1+DESI-DR1 & $0.3<z<0.4$ & $0.51 \pm 0.18$ & $3<R<80$ & \textit{Current}   \\
      25 & HSC-Y3+DESI-DR1 & $0.3<z<0.4$ & $0.56 \pm 0.17$ & $3<R<80$ & \textit{Current}   \\
       & & & & & \\    
      26 & KiDS-1000+BOSS-DR12+2dFLenS & $0.4<z<0.5$ & $0.33 \pm 0.06$ & $3<R<100$ & \cite{blake_testing_2020}  \\
      27 & KiDS-1000+BOSS-DR12 & $0.4<z<0.5$ & $0.27 \pm 0.08$ & $3<R<50$ &  \textit{Current} \\
      28 & DES-Y3+BOSS-DR12 & $0.4<z<0.5$ & $0.35 \pm 0.06$ & $3<R<50$ & \textit{Current}   \\
      29 & HSC-Y1+BOSS-DR12 & $0.4<z<0.5$ & $0.37 \pm 0.09$ & $3<R<50$ &  \textit{Current}  \\
       & & & & & \\
      30 & KiDS-1000+DESI-DR1 & $0.4<z<0.6$ & $0.34 \pm 0.07$ & $3<R<80$ & \textit{Current}   \\
      31 & DES-Y3+DESI-DR1 & $0.4<z<0.6$ & $0.25 \pm 0.06$ & $3<R<80$ & \textit{Current}   \\
      32 & HSC-Y1+DESI-DR1 & $0.4<z<0.6$ & $0.36 \pm 0.08$ & $3<R<80$ & \textit{Current}   \\
      33 & HSC-Y3+DESI-DR1 & $0.4<z<0.6$ & $0.38 \pm 0.07$ & $3<R<80$ & \textit{Current}   \\
       & & & & & \\
      34 & RCSLenS+CFHTLenS+WGZHiZ+CMASS &$0.43<z<0.7$ & $0.31 \pm 0.06$ & $3<R<50$ & \cite{blake_rcslens_2016}  \\
      35 & CFHTLenS+CMASS & $0.43<z<0.7$ & $0.42 \pm 0.056$ & $5<R<26$ & \cite{alam_testing_2017}  \\
      36 & KiDS-450+CMASS+2dFHIZ & $0.43<z<0.7$ & $0.26 \pm 0.07$ & $5<R<60$ &  \cite{amon_kids2dflensgama_2018} \\
      37 & RCSLenS+CFHTLenS+WGZHiZ+CMASS  & $0.43<z<0.7$ & $0.30 \pm 0.07$ & $10<R<50$ & \cite{blake_rcslens_2016}  \\ 
      38 & CFHTLenS+CS82+CMASS & $0.43<z<0.7$ & $0.43 \pm 0.11$ & $10<R<60$ & \cite{jullo_testing_2019}  \\ 
      39 & SDSS+CMASS & $0.45<z<0.7$ & $0.39 \pm 0.05 $ & $25<R<150$ &  \cite{singh_probing_2019} \\
      40 & CFHTLenS+VIPERS  & $0.5<z<0.7$ & $0.16 \pm 0.09$ & $3<R<20$ & \cite{de_la_torre_vimos_2017} \\
       & & & & & \\
      41 & KiDS-1000+BOSS-DR12+2dFLenS & $0.5<z<0.6$ & $0.38 \pm 0.07$ & $3<R<100$ & \cite{blake_testing_2020}  \\
      42 & KiDS-1000+BOSS-DR12 & $0.5<z<0.6$ & $0.46 \pm 0.09$ & $3<R<50$ &  \textit{Current} \\
      43 & DES-Y3+BOSS-DR12 & $0.5<z<0.6$ & $0.28 \pm 0.06$ & $3<R<50$ & \textit{Current}   \\
      44 & HSC-Y1+BOSS-DR12 & $0.5<z<0.6$ & $0.51 \pm 0.09$ & $3<R<50$ &  \textit{Current}  \\
      & & & & & \\
      45 & KiDS-1000+BOSS-DR12+2dFLenS & $0.6<z<0.7$ & $0.34 \pm 0.08$ & $3<R<100$ & \cite{blake_testing_2020}  \\
      46 & KiDS-1000+BOSS-DR12 & $0.6<z<0.7$ & $0.33 \pm 0.10$ & $3<R<50$ &  \textit{Current} \\
      47 & DES-Y3+BOSS-DR12 & $0.6<z<0.7$ & $0.39 \pm 0.08$ & $3<R<50$ & \textit{Current}   \\
      48 & HSC-Y1+BOSS-DR12 & $0.6<z<0.7$ & $0.52 \pm 0.11$ & $3<R<50$ &  \textit{Current}  \\
      & & & & & \\
      49 & KiDS-1000+DESI-DR1 & $0.6<z<0.8$ & $0.29 \pm 0.06$ & $3<R<80$ & \textit{Current}   \\
      50 & DES-Y3+DESI-DR1 & $0.6<z<0.8$ & $0.39 \pm 0.07$ & $3<R<80$ & \textit{Current} \\
      51 & HSC-Y1+DESI-DR1 & $0.6<z<0.8$ & $0.36 \pm 0.07$ & $3<R<80$ & \textit{Current} \\
      52 & HSC-Y3+DESI-DR1 & $0.6<z<0.8$ & $0.35 \pm 0.06$ & $3<R<80$ & \textit{Current} \\
      & & & & & \\      
      53 & CFHTLenS+VIPERS  & $0.7<z<1.2$ & $0.09 \pm 0.07  $ & $3<R<20$ & \cite{de_la_torre_vimos_2017}  \\
      & & & & & \\
      54 & KiDS-1000+DESI-DR1 & $0.8<z<1.1$ & $0.35 \pm 0.15$ & $3<R<80$ & \textit{Current}   \\
      55 & DES-Y3+DESI-DR1 & $0.8<z<1.1$ & $0.22 \pm 0.12$ & $3<R<80$ & \textit{Current} \\
      56 & HSC-Y1+DESI-DR1 & $0.8<z<1.1$ & $0.35 \pm 0.09$ & $3<R<80$ & \textit{Current} \\ 
      57 & HSC-Y3+DESI-DR1 & $0.8<z<1.1$ & $0.35 \pm 0.07$ & $3<R<80$ & \textit{Current} \\
      & & & & & \\
\end{longtable}
\endgroup

\end{document}